%% file: TOP-17-003_temp.tex
\pdfoutput=1

\documentclass[11pt,twoside,a4paper,cmspaper,final,collab]{cms-tdr}
\def\svnVersion{466791P}\def\cmsCernNoTag{CERN-EP-2017-309}\def\cmsCernDate{\today}\def\cmsMessage{Published in the Journal of High Energy Physics as \href{http://dx.doi.org/10.1007/JHEP06(2018)102}{\doi{10.1007/JHEP06(2018)102.}}}

\begin{document}\cmsNoteHeader{TOP-17-003}

\hyphenation{had-ron-i-za-tion}
\hyphenation{cal-or-i-me-ter}
\hyphenation{de-vices}
\RCS$Revision: 466684 $
\RCS$HeadURL: svn+ssh://svn.cern.ch/reps/tdr2/papers/TOP-17-003/trunk/TOP-17-003.tex $
\RCS$Id: TOP-17-003.tex 466684 2018-06-29 07:31:12Z kskovpen $

\providecommand{\FEYNRULES}{\textsc{FeynRules}\xspace}

\cmsNoteHeader{TOP-17-003}
\title{Search for the flavor-changing neutral current interactions of the top
quark and the Higgs boson which decays into a pair of b quarks at $\sqrt{s}=13\TeV$}

\date{\today}

\abstract{
A search for flavor-changing neutral currents (FCNC) in events with
the top quark and the Higgs boson is presented. The Higgs boson
decay to a pair of b quarks is considered.
The data sample corresponds to an integrated luminosity of
$35.9\fbinv$
recorded by the CMS experiment at the LHC in proton-proton collisions
at $\sqrt{s}=13\TeV$.
Two channels are considered: single top quark FCNC production in
association with the Higgs boson ($\Pp\Pp \to \PQt\PH$),
and top quark pair production with FCNC decay of the top quark ($\PQt \to \PQq\PH$).
Final states with one isolated lepton and at least three
reconstructed jets, among which at least two are associated with $\PQb$
quarks, are studied.
No significant deviation is observed
from the predicted background. Observed (expected) upper limits at 95\% confidence level are
set on the branching fractions of top quark decays, $\mathcal{B}(\PQt \to \PQu\PH) <0.47\%\,(0.34\%)$
and $\mathcal{B}(\PQt \to \PQc\PH) <0.47\%\,(0.44\%)$, assuming a
single nonzero FCNC coupling.}

\hypersetup{%
pdfauthor={CMS Collaboration},%
pdftitle={Search for the flavor-changing neutral current interactions of the top
quark and the Higgs boson which decays into a pair of b
quarks at sqrt(s) = 13 TeV},%
pdfsubject={CMS},%
pdfkeywords={CMS, physics, Higgs, top, FCNC}}

\maketitle

\section{Introduction}
\begin{figure}[hb]
  \centering
    \includegraphics[width=0.4\textwidth]{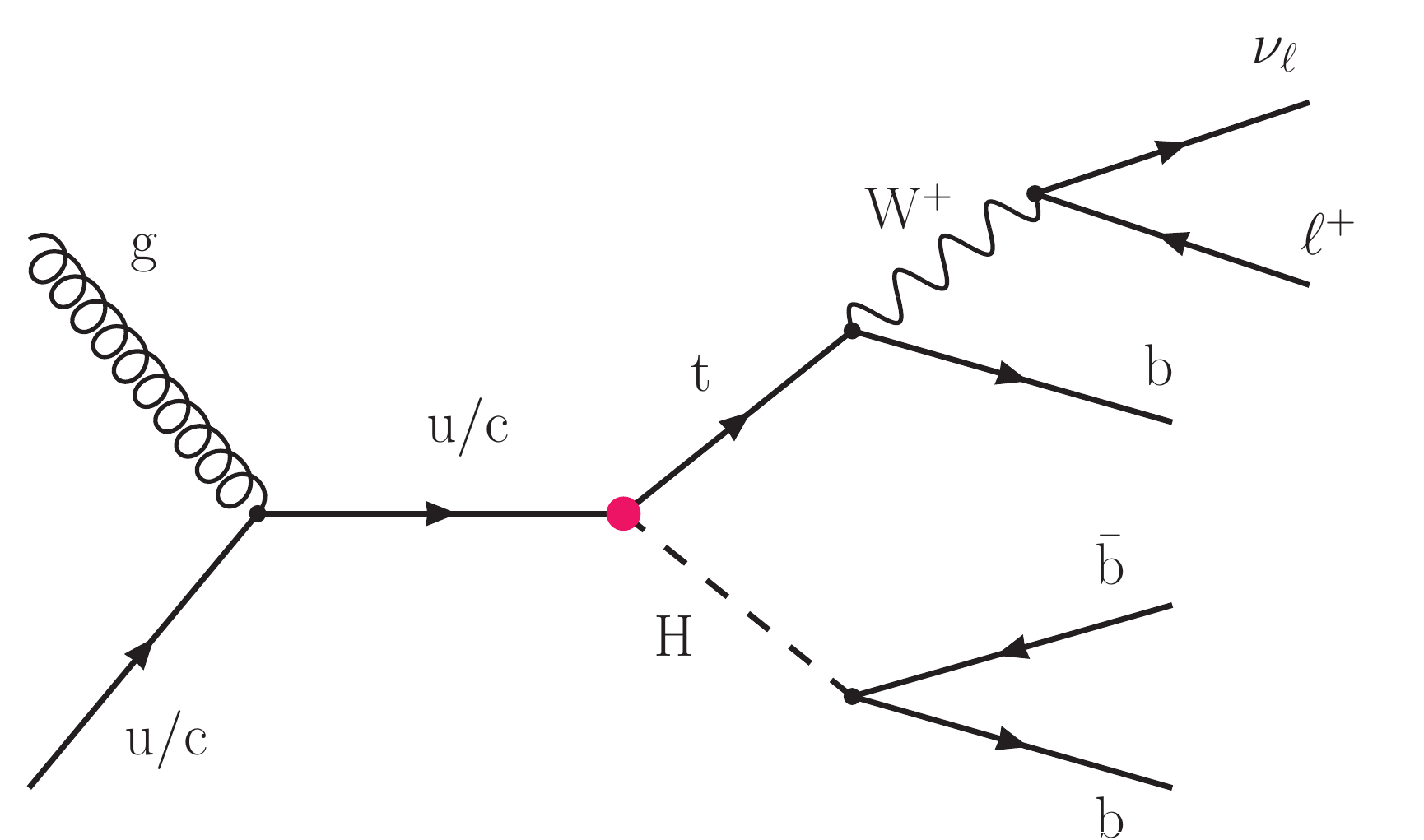}
    \includegraphics[width=0.4\textwidth]{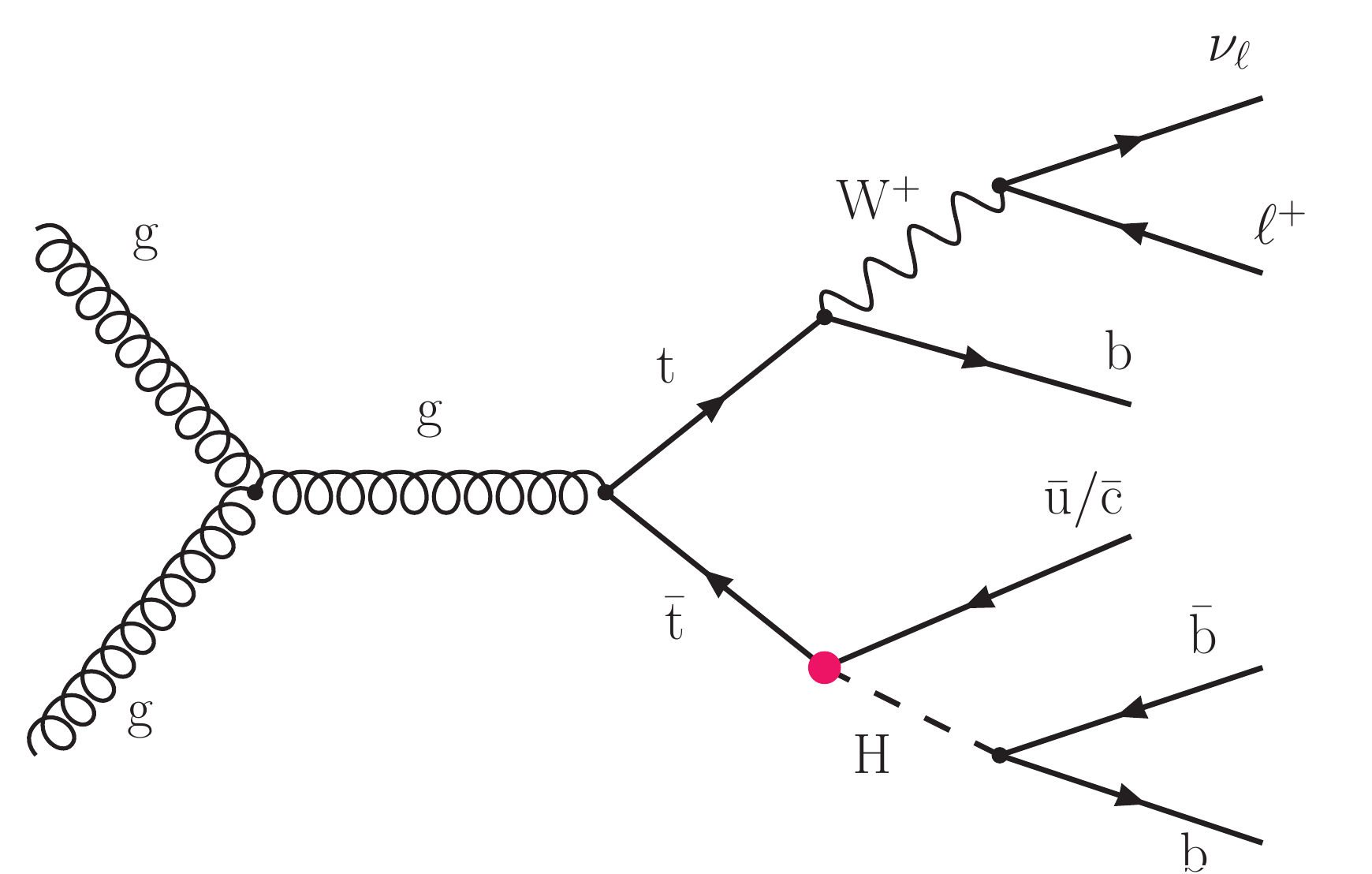}
    \caption{Representative Feynman diagrams for FCNC $\PQt\PH$ processes: associated
    production of the top quark with the Higgs boson (left), and FCNC decay of
    the top antiquark in \ttbar events (right). The FCNC vertex is indicated by the bullet.}
    \label{fig:diag}
\end{figure}

A recently discovered fundamental particle
has properties that are consistent
with the standard model (SM) predictions for the Higgs boson, $\PH$~\cite{HIG1,HIG2,HIG3,HIG4}.
In the SM, flavor-changing neutral currents
(FCNC) are forbidden at tree level and are strongly suppressed in loop
corrections by the Glashow--Iliopoulos--Maiani (GIM) mechanism~\cite{GIM} with the SM branching fraction of
$\PQt \to \PQq\PH$ predicted to be $\mathcal{O}(10^{-15})$~\cite{FCNC1,FCNC2,FCNC3}.
Several extensions of the SM incorporate significantly enhanced FCNC behavior
that can be directly probed at the CERN LHC~\cite{FCNC3,FCNCREV}. The FCNC processes that correspond
to $\PQt\PH$ interactions are described by the following effective Lagrangian:
\begin{equation}
\mathcal{L} = \sum_{\PQq=\PQu,\PQc}
\frac{g}{\sqrt{2}} \PAQt\, \kappa_{\PH\PQq\PQt}\,
\big(f_{\PH\PQq}^{\mathrm{L}} P_{\mathrm{L}} +
f_{\PH\PQq}^{\mathrm{R}} P_{\mathrm{R}}\big)\,\PQq\, \PH + \text{h.c.},\label{lagrangian}
\end{equation}
where $g$ is the weak coupling constant, $P_{\mathrm{L}}$ and $P_{\mathrm{R}}$ are chirality projectors in spin space,
$\kappa_{\PH\PQq\PQt}$ is the effective coupling,
$f_{\PH\PQq}^{\mathrm{L}}$ and $f_{\PH\PQq}^{\mathrm{R}}$ are left- and right-handed complex chiral
parameters with a unitarity constraint of $|f^{\mathrm{L}}_{\PH\PQq}|^2 +
|f^{\mathrm{R}}_{\PH\PQq}|^2 = 1$.
The $\PQt\PH$ FCNC interaction is studied in this analysis in two channels: the associated
production of a single top quark with the Higgs boson (ST), as well as in FCNC decays of top quarks in \ttbar
semileptonic events (TT).
In this analysis, $\PH \to \bbbar$ decays are considered.
This is the first time that the analysis of
the ST mode is presented. Representative Feynman diagrams of the studied processes are
shown in Fig.~\ref{fig:diag}.

Earlier analyses by the ATLAS~\cite{TopHiggsATLAS,TopHiggsATLAS13TeV} and
CMS~\cite{TopHiggsCMS} Collaborations have probed $\kappa_{\PH\PQq\PQt}$ in top quark decays in
\ttbar events. The ATLAS search at center-of-mass energy of 13\TeV investigated the
$\PQt \to \PQq\PH$ decay with the Higgs boson decaying to two photons
to set observed (expected) upper
limits at $95\%$ confidence level (CL) on the branching fractions $\mathcal{B}(\PQt \to \PQu\PH)$
and $\mathcal{B}(\PQt \to \PQc\PH)$ of 0.24\% (0.17\%) and 0.22\% (0.16\%),
respectively~\cite{TopHiggsATLAS13TeV}.
The CMS analysis at $\sqrt{s}=8\TeV$ utilized the Higgs boson decays into either boson or fermion pairs
to set observed (expected) upper limits of 0.55\% (0.40\%) and 0.40\%
(0.43\%) on $\mathcal{B}(\PQt \to \PQu\PH)$ and
$\mathcal{B}(\PQt \to \PQc\PH)$, respectively~\cite{TopHiggsCMS}.

For the signal processes, we consider the cross section times branching
fraction with a specific signature for single top quark $\PQt(\to
\ell^+ \PGn\PQb)\PH(\to \bbbar)$
and pair production $\PQt(\to \ell^+ \PGn\PQb)
\PAQt(\to \PAQu/\PAQc\PH(\to \bbbar) )$, with $\ell$ =
$\mathrm{e}$, $\mu$, or $\tau$.
The analysis also considers the charge-conjugate process.
The predicted cross section at 13\TeV for single top quark and antiquark FCNC
production in association with the Higgs boson under the assumption of
coupling strengths $\kappa_{\PH\PQu\PQt}=1,
\kappa_{\PH\PQc\PQt}=0$ ($\kappa_{\PH\PQc\PQt}=1, \kappa_{\PH\PQu\PQt}=0$) is
13.8\,(1.90)\unit{pb}, where the cross section calculation is based on the
leading order (LO) set of NNPDF 2.3 parton distribution functions (PDFs)~\cite{TT5}.
In the case of the production of \ttbar semileptonic events with top quark FCNC decay, the
predicted cross section is 37.0\unit{pb} and is independent of the
type of the coupling.
By exploiting a simultaneous analysis of both the TT and ST processes, an
improved sensitivity to $\kappa_{\PH\PQu\PQt}$ can be achieved, as the
ST production via the up quark is enhanced by the proton PDFs.

This analysis uses data that correspond to an integrated luminosity
of 35.9\fbinv~\cite{lumi} recorded in 2016 by the
CMS experiment at the LHC in proton-proton (pp) collisions
at $\sqrt{s}=13\TeV$. Events with exactly one isolated lepton
(electron or muon) and with at least three jets, among which at least
two are associated with $\PQb$ quarks, are considered.

\section{The CMS detector}

The central feature of the CMS detector is a superconducting solenoid
of 6\unit{m} internal diameter, providing a magnetic field of
3.8\unit{T}. Within the solenoid volume are a silicon pixel and strip
tracker, a lead tungstate crystal electromagnetic calorimeter (ECAL),
and a brass and scintillator hadron calorimeter (HCAL), each composed
of a barrel and two endcap sections. Forward calorimeters extend the
pseudorapidity coverage provided by the barrel and endcap detectors.
Muons are measured in gas-ionization detectors embedded in the steel
flux-return yoke outside the solenoid. A more detailed description of the CMS detector, together with a
definition of the coordinate system used and the relevant kinematic
variables, can be found in Ref.~\cite{Chatrchyan:2008zzk}.

\section{Monte Carlo simulation}

The generation of simulated signal events is done at LO with
\MGvATNLO 2.3.3 \cite{MG1,MG2}. Up to two additional partons are simulated
by the Monte Carlo (MC) generator in the initial hard process for the top quark pair production mode.
The MLM~\cite{MLM} matching scheme is used to match additional partons in
the matrix-element calculations to the parton-shower description.
No additional partons are included in the generation of events for the single
top quark production process, as such inclusion would contain
contributions from the top quark pair production process.
A systematic variation in the normalization of the single top production process by 10\% is considered in order to
account for the differences in the generation of additional radiation of the two signal production modes.
The Lagrangian terms from~Eq.~(\ref{lagrangian}) are
implemented by means of the \FEYNRULES
package~\cite{feynrules} using the universal \FEYNRULES output
format~\cite{ufo}. The complex chiral parameters are fixed to
$f_{\mathrm{Hq}}^{\mathrm{R}}=1$ and $f_{\mathrm{Hq}}^{\mathrm{L}}=0$.

The SM top quark pair production is the dominant background process and
is simulated to next-to-leading order (NLO) using \POWHEG
v2~\cite{POWHEG1,POWHEG2,POWHEG3,POWHEG4}.
The predicted cross section for this process is
832~${}^{+20}_{-29}\,\text{(scale)}\pm35$(PDF)\unit{pb},
as calculated with the $\textsc{Top++}$ 2.0 program at next-to-next-to-leading
order (NNLO), including soft-gluon resummation to
next-to-next-to-leading-log order (see Ref. \cite{TT1} and references therein),
and assuming a top quark mass of $m_{\PQt} = 172.5$\GeV.
Two systematic uncertainties that are shown in the
prediction are considered. These are independent variations of the
factorization and renormalization scales, $\mu_{\mathrm{F}}$ and
$\mu_{\mathrm{R}}$, and variations of the
PDF and $\alpha_{s}$.

Single top quark production in the $t$ channel is simulated with
\POWHEG v2 in the four-flavour scheme,
while events for single top quark production in association with W
bosons are generated with \POWHEG v1 in the five-flavour scheme (5FS).
The predicted NLO cross sections are $217^{+9}_{-8}$~\cite{tchan1,tchan2} and $71.7\pm3.9$\unit{pb}~\cite{Wt}, respectively.
Single top quark production in the $s$ channel is done at NLO with the
\MGvATNLO generator in 5FS with a predicted cross section
of $10.3\pm0.4$\unit{pb}. The uncertainties in the quoted cross sections
correspond to independent variations of $\mu_{\mathrm{F}}$ and
$\mu_{\mathrm{R}}$, as well as to variations of the PDF and $\alpha_{s}$.
Small contributions to the overall predicted
background arise from several additional processes: W boson production and the
associated production of \ttbar with W and Z, both generated with
\MGvATNLO, and from Drell--Yan and the associated production of \ttbar with a
Higgs boson generated with the \MGvATNLO and \POWHEG v1~\cite{POWHEGDY}, respectively.

In the simulation of signal and background
processes, the initial- and final-state radiation (ISR and FSR), as
well as the fragmentation and hadronization of
quarks, are modeled using \PYTHIA~8.212~\cite{PYTHIA2} with the underlying event tune
CUETP8M1~\cite{CUETP8M1}. For \ttbar generation, the first emission is
done at the matrix-element level with \POWHEG~v2. Generation of \ttbar and
single top quark production in the $t$ channel uses the underlying event tune CUETP8M2T4~\cite{CMS-PAS-TOP-16-021}.
In the generation of all background processes the NNPDF3.0
PDF~\cite{NNPDF} set is used.

The detector response is simulated using \GEANTfour~v9.4~\cite{Geant4}.
In order to model the effect of multiple interactions per event
crossing (pileup), generated minimum bias events were added to the
simulated data. The number of extra multiple interactions were matched
to agree with the rate observed in data.
The number of pileup interactions in data is estimated from the measured bunch-to-bunch instantaneous
luminosity and the total inelastic cross section (69.2\unit{mb}) \cite{lumi}.

\section{Event selection}

The particle-flow (PF) algorithm~\cite{CMS-PFT} reconstructs and identifies each
individual particle with an optimized combination of information from
the various elements of the CMS detector. The energy of photons is
directly obtained from the ECAL measurement, corrected for
zero-suppression effects. The energy of electrons is determined from a
combination of the electron momentum at the primary interaction vertex
as determined by the tracker, the energy of the corresponding ECAL
cluster, and the energy sum of all bremsstrahlung photons spatially
compatible with originating from the electron track. The momentum of
muons is obtained from the curvature of the corresponding track. The
energy of charged hadrons is determined from a combination of their
momentum measured in the tracker and the matching ECAL and HCAL energy
deposits, corrected for zero-suppression effects and for the response
function of the calorimeters to hadronic showers. Finally, the energy
of neutral hadrons is obtained from the corresponding corrected ECAL
and HCAL energy.

Jets are reconstructed by clustering PF candidates using the anti-\kt
algorithm~\cite{Cacciari:2008gp, Cacciari:2011ma} with a distance
parameter of 0.4. The jet momentum is determined as the vectorial sum of all particle
momenta in the jet, and is found from simulation to be within 5 to
10\% of the true momentum over the whole transverse momentum (\pt) spectrum and detector
acceptance~\cite{Chatrchyan:2011ds}. An offset correction is applied to jet energies to take
into account the contribution from pileup.
Jet energy
corrections are derived from simulation and are confirmed with in
situ measurements of the energy balance in dijet, multijet,
$\gamma$+jet, and leptonic $\mathrm{Z}$+jet events. Additional selection criteria are applied to each event to
remove spurious jet-like features originating from isolated noise
patterns in certain HCAL regions.
The missing transverse momentum ($\ptvecmiss$) in an event is defined as the magnitude of the transverse projection of the vector sum of
the momenta of all reconstructed PF candidates in an event.

The reconstructed vertex with the largest value of summed
physics-object $\pt^2$ is taken to be the primary $\Pp\Pp$ interaction
vertex. The physics objects are the jets, clustered using the jet
finding algorithm~\cite{Cacciari:2008gp,Cacciari:2011ma} with the
tracks assigned to the vertex as inputs, and the associated $\ptvecmiss$,
taken as the negative vector \pt sum of those jets, to represent the neutral particles.

This analysis selects events with exactly one isolated lepton (electron
or muon).
Events with one electron (muon) are recorded using a trigger that
required at least one electron (muon) with $\pt >32\,(24)$\GeV
selected within the detector acceptance ($|\eta| <2.1$).
Electron (muon) candidates are selected offline with $|\eta| <2.1$
with $\pt >35\,(30)$\GeV. Electrons that are reconstructed in the
transition region between the barrel and endcap regions of the ECAL, $1.44 < |\eta| < 1.57$, are removed.
Leptons are required to be isolated in terms of a relative isolation
variable, $I_{\text{rel}}$.
This variable is defined as the ratio of the scalar \pt sum of photons,
charged hadrons, and neutral hadrons within a cone of angular radius
$\Delta R = \sqrt{\smash[b]{(\Delta\eta)^2+(\Delta\phi)^2}} = 0.3\,(0.4)$ of the
reconstructed lepton candidate, where $\phi$ is azimuthal angle in
radians, to the lepton \pt.
This isolation variable only includes the charged hadrons that emerge from the same vertex as the
selected lepton and is corrected for energy deposits from neutral
particles produced in pileup interactions. For electron (muon)
candidates, $I_{\text{rel}}$ must be less than 0.06 (0.15).
In order to suppress background processes with multilepton final
states, events with additional leptons passing the looser isolation
requirement of $I_{\text{rel}} <0.25$ and $\pt > 10$\GeV are rejected.

At least three jets are required to be present in the event with
$\pt >30$\GeV and $|\eta| <2.4$.
As signal events contain three b quarks produced in the final state
at the tree level, we require at least two jets are identified
as b quark jets by the combined secondary vertex
v2 (CSVv2) b tagging algorithm~\cite{CMS-PAS-BTV-15-001}.
This requirement corresponds to the selection of jets with the CSVv2
discriminant value greater than 0.85, and provides a b jet efficiency
of {$\approx$}70\%, with a misidentification rate for c jets and jets
originating from light quarks and gluons of {$\approx$}10\% and
{$\approx$}1\%, respectively.

\section{Event reconstruction and multivariate analysis}

In order to optimize sensitivity to the
signal event selection, events are split into five categories based on the total number of
reconstructed jets and on the number of b-tagged jets. Categories with exactly three jets of which  two or
three are identified as b jets are denoted as b2j3 and b3j3,
respectively.
Similarly, categories with at least four jets
of which two, three, or four are identified as b jets are specified as
b2j4, b3j4, and b4j4, respectively
The longitudinal momentum of the neutrino is determined by assigning
$\ptvecmiss$ to the neutrino, and constraining the
$\ell\nu$ mass to the known mass of the W boson.
With the use of the energy and momenta of all particles, a full kinematic reconstruction of the event
is performed for several signal and background hypotheses: ST, TT, and
background \ttbar events, where one of the top quarks decays
semileptonically, and the other one hadronically.
The reconstruction is performed
for all possible permutations of the b-tagged jets
to be associated with the decay products of the Higgs boson or the top
quark, and both solutions for the longitudinal momentum of the neutrino are considered.
The reconstructed kinematic variables for each permutation are then fed
into a multivariate analysis that uses a boosted decision tree
(BDT)~\cite{BDT}
approach, as implemented in the toolkit for multivariate
analysis TMVA~\cite{TMVA}.
The input BDT variables that are used for the ST and TT
hypotheses correspond to the reconstructed invariant mass of two b
jets associated with the Higgs boson decay, the reconstructed
invariant mass of a b jet ($m_{\mathrm{b\overline{b}}}$), lepton and neutrino associated with the top
quark decay ($m(\mathrm{t}^{\ell})$), its transverse momentum
($\pt(\mathrm{t}^{\ell}$)), $\Delta R$ between the reconstructed Higgs
boson and the top quark. In case of the hypothesis of the background
$t\bar{t}$ events the following variables are used:
$m(\mathrm{t}^{\ell})$, $m(\mathrm{t}^{\mathrm{h}})$,
$\Delta R(\mathrm{t}^{\ell},\mathrm{t}^{\mathrm{h}})$, and
$\pt(\mathrm{t}^{\ell})$, where $\mathrm{t}^{\mathrm{h}}$
corresponds to the reconstructed top quark hadronic decay from one b-tagged and two
non b-tagged jets.
The BDT classifier is trained to distinguish the correct from the
wrong b jet assignments. The training and validation of the BDT is
performed on statistically independent simulated samples.
All reconstructed b jets in the event are considered, and the
permutation with the highest BDT score is chosen as the correct one.
The measured
algorithm efficiency for correct assignment of the b-tagged jets to the
jets reconstructed at generator level after applying the analysis
selection criteria is {$\approx$}75\%.

\begin{figure*}
  \centering
    \includegraphics[width=0.45\textwidth]{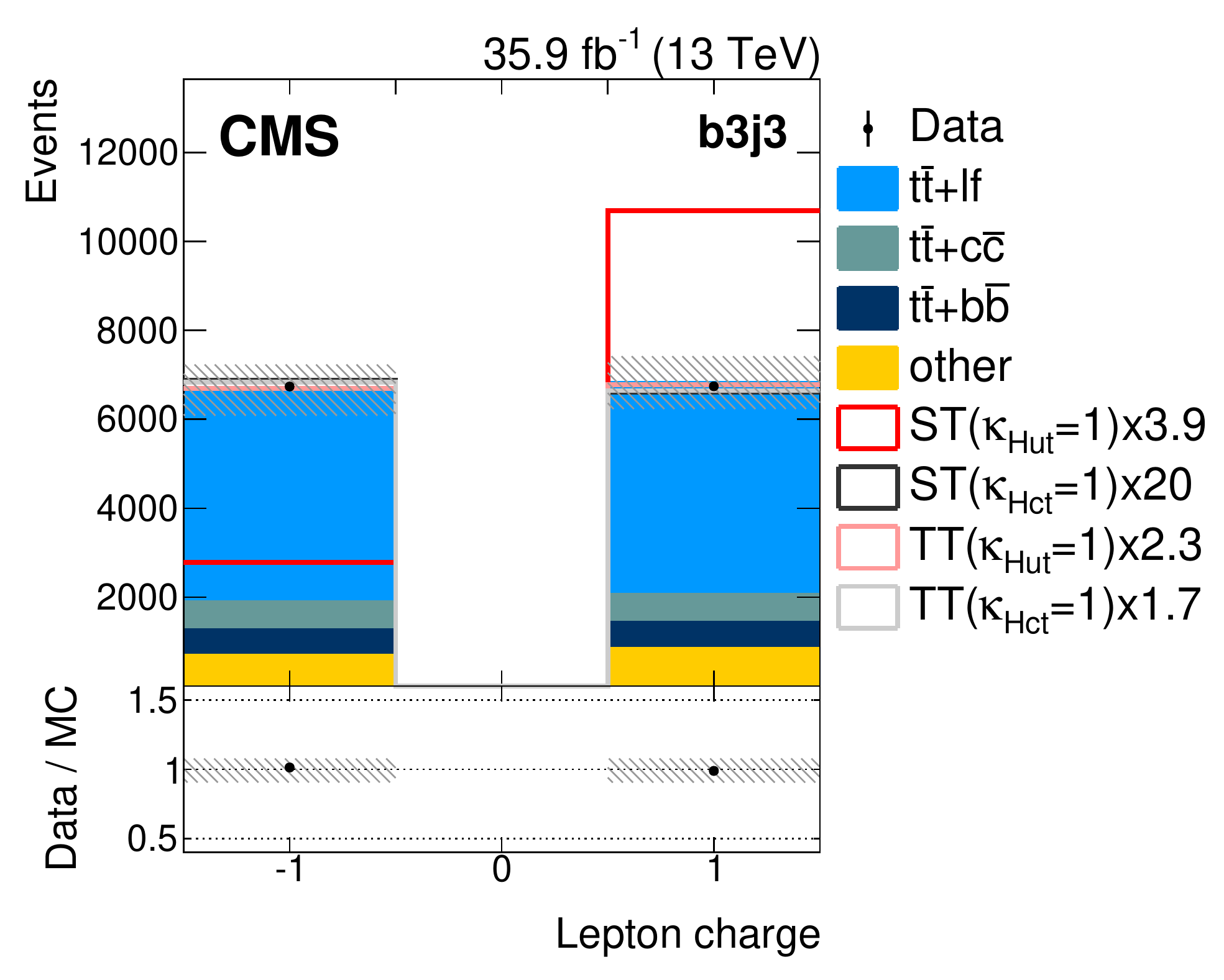}
    \includegraphics[width=0.45\textwidth]{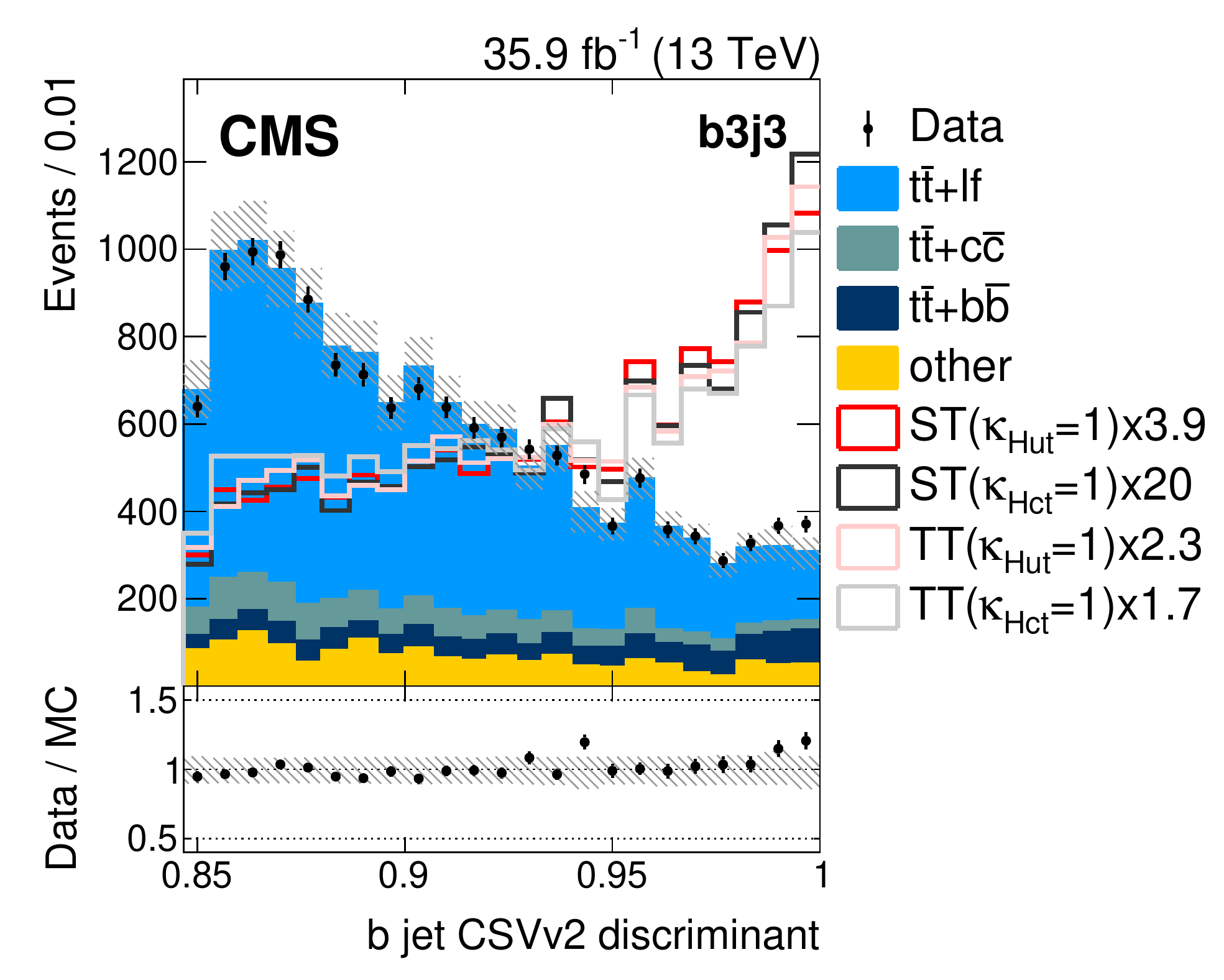}
    \includegraphics[width=0.45\textwidth]{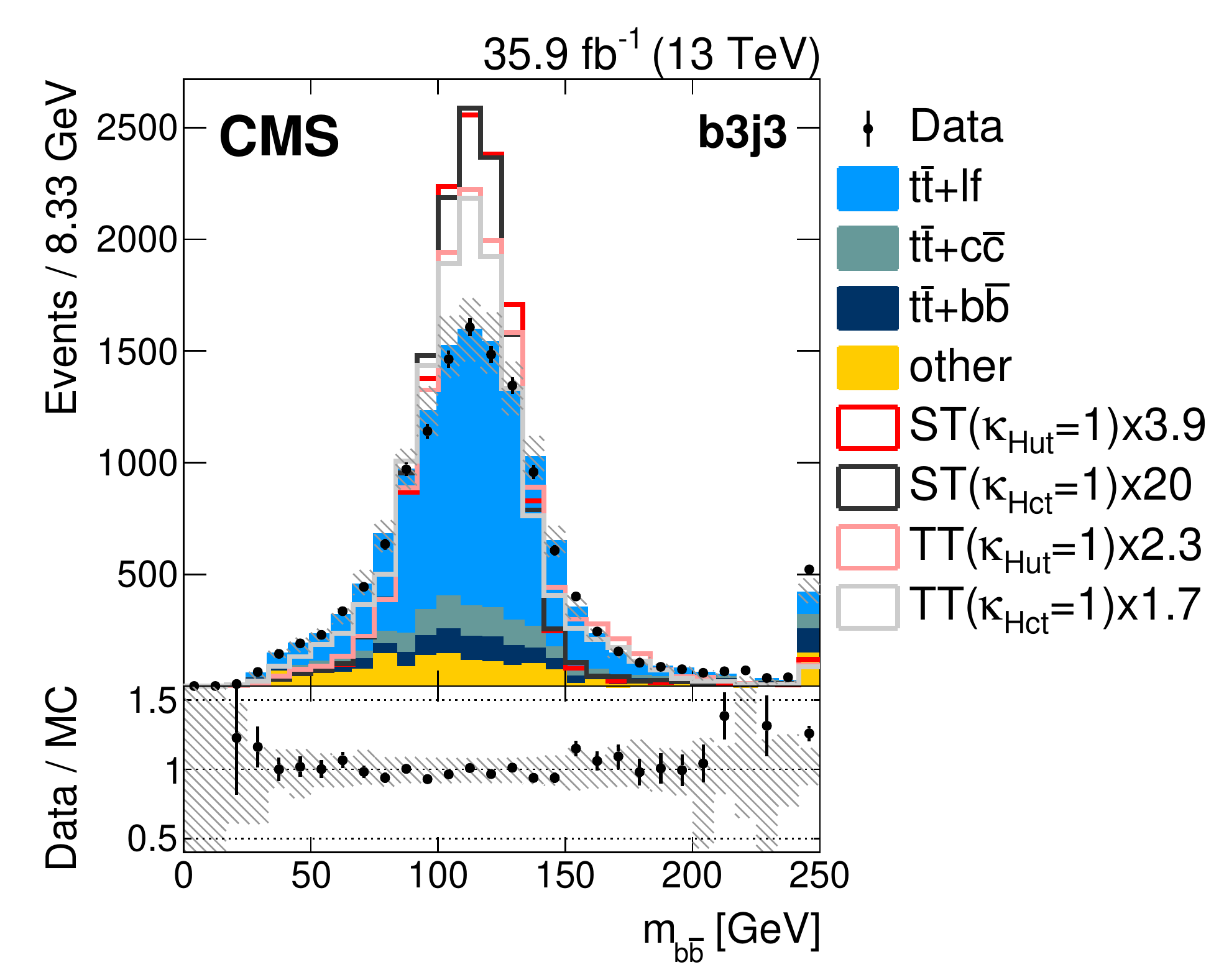}
    \includegraphics[width=0.45\textwidth]{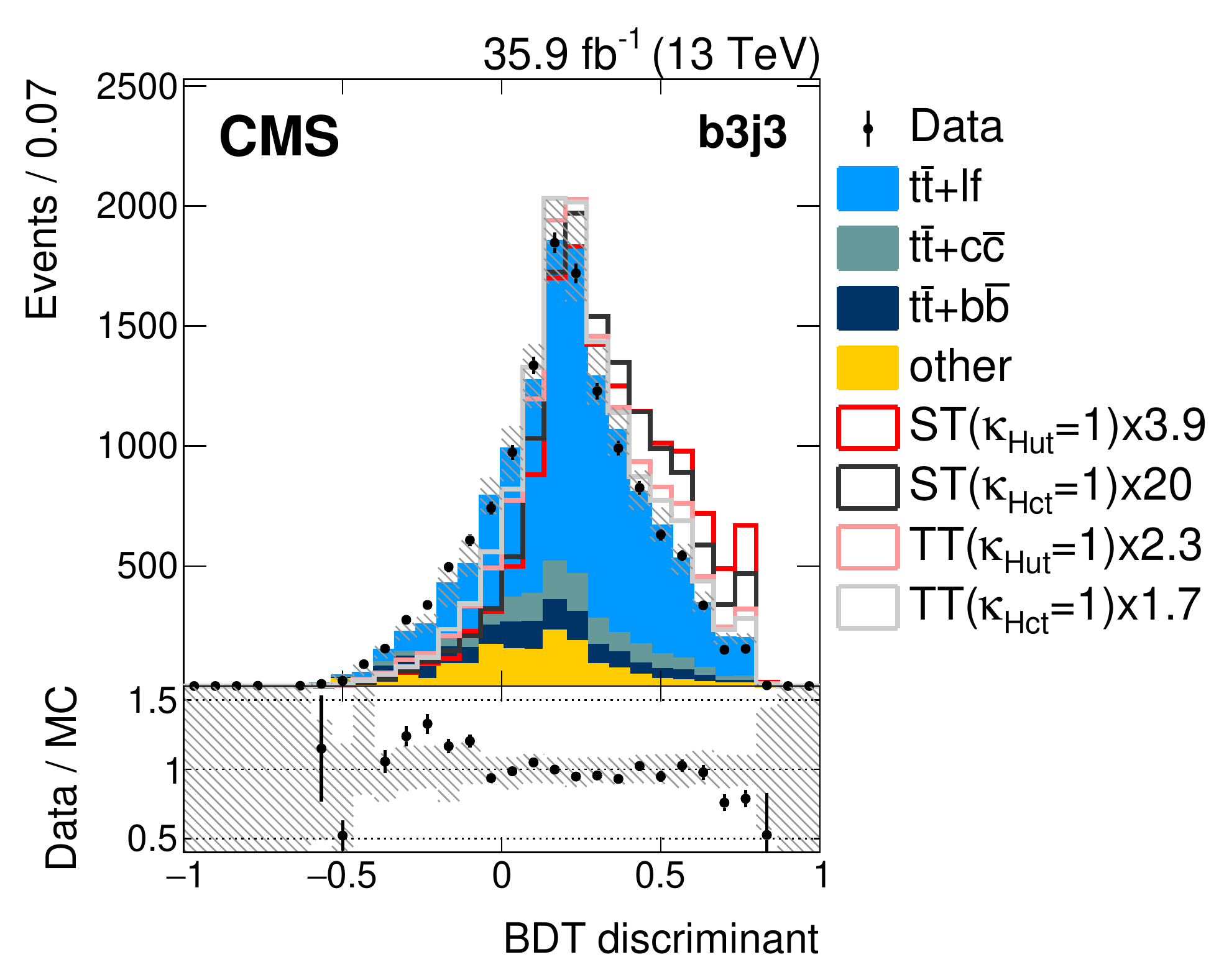}
    \caption{Comparison between data and simulation for the most
    discriminating BDT input variables in the b3j3 category: lepton
    charge (upper left), CSVv2 discriminant
    value for one of the reconstructed b jets assigned to Higgs boson decay
    (upper right), reconstructed invariant mass of two b jets
    associated with the Higgs boson decay (lower left),
    and the maximum BDT discriminant value from the b jet assignment
    procedure (lower right).
    The last bin in the distribution for the reconstructed mass of
    the Higgs boson includes the overflows.
    The shaded area corresponds to the total uncertainty
    in the predicted background.
    The data-to-simulation ratio is also shown. The distributions for
    the signal processes are normalized to the total number of events in
    the predicted background to ease the comparison
    of the shapes of the distributions.}
    \label{fig:BDTvar}
\end{figure*}

Kinematic variables from the event reconstruction are used to
construct several BDTs to suppress backgrounds.
The BDTs are trained for each jet multiplicity category to identify signal events
that are generated either for $\kappa_{\mathrm{Hut}}$ (Hut) or
$\kappa_{\mathrm{Hct}}$ (Hct) coupling against the sum of all
background events. Separate trainings of the BDT for Hut and Hct are
done in order to take into account the differences in kinematic properties
of the reconstructed objects in the ST production mode, as well as the
differences in the measured b tagging efficiencies for a charm and an up quark
in the TT production channel.
The most important variables that discriminate between signal
and background events are: the charge of the lepton (considered only for
the BDT that uses Hut signal events), the CSVv2 discriminant
value of the b jet with the lowest \pt from the Higgs boson decay,
$m_{\mathrm{b\overline{b}}}$, and the output
discriminant value of the BDT used in the b jet assignment procedure.
Distributions for these variables in data and MC simulation in the b3j3 category are presented in
Fig.~\ref{fig:BDTvar}.
The b4j4 category is not considered for Hut due to negligible improvement in the
final sensitivity.

The simulated \ttbar background events are
split into subcategories defined by the flavor of additional
particle-level jets produced in association with the top quark pair.
These classes are referred to as
\ttbar{}+\bbbar,
\ttbar{}+\ccbar,
and \ttbar{}+lf,
(where lf stands for light flavor).
The \ttbar{}+lf category includes events
where no additional pair of b or c jets occurs. The other background
processes are summed up and shown together in the prediction.

\begin{figure}[htb]
  \centering
    \includegraphics[width=0.45\textwidth]{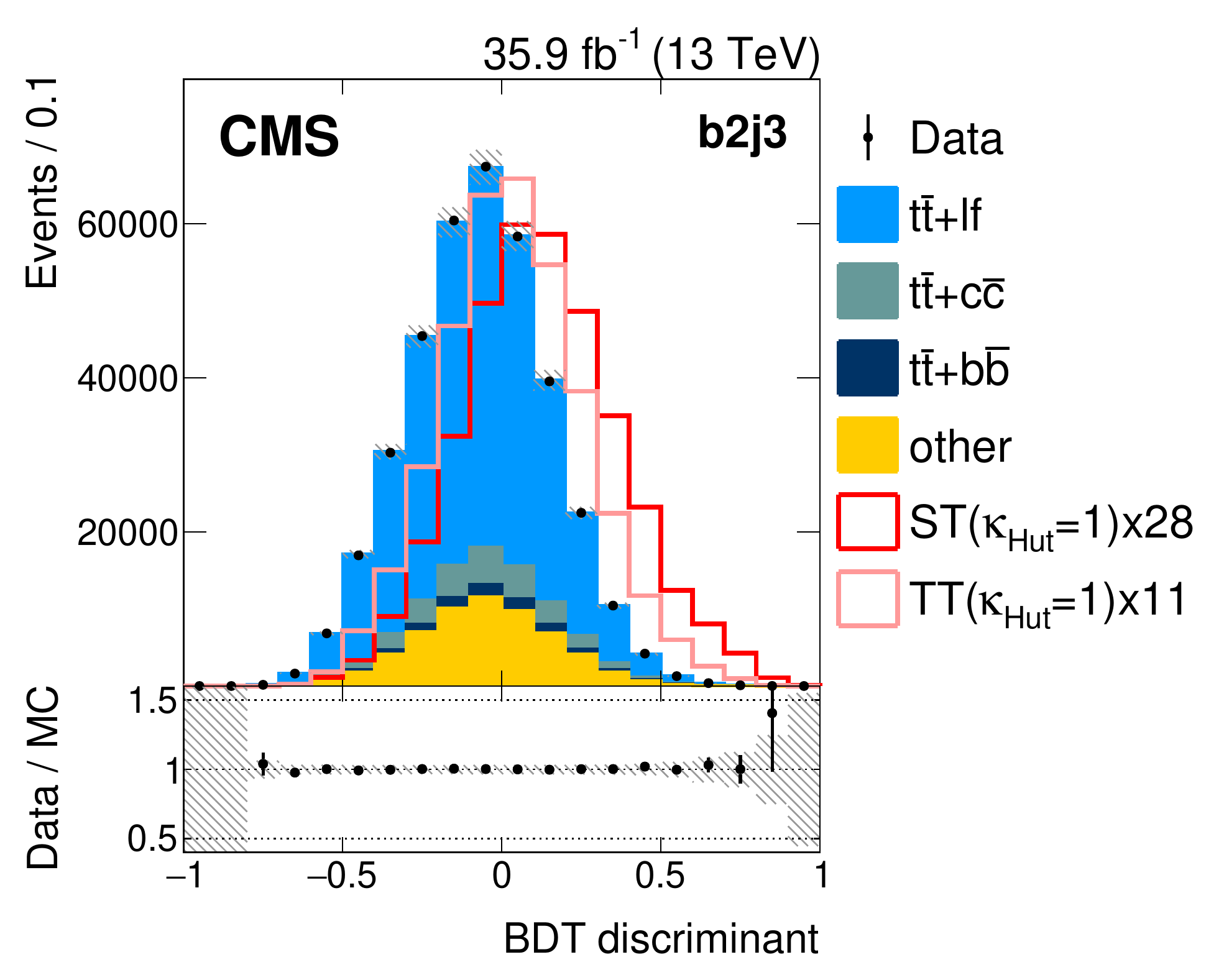}
    \includegraphics[width=0.45\textwidth]{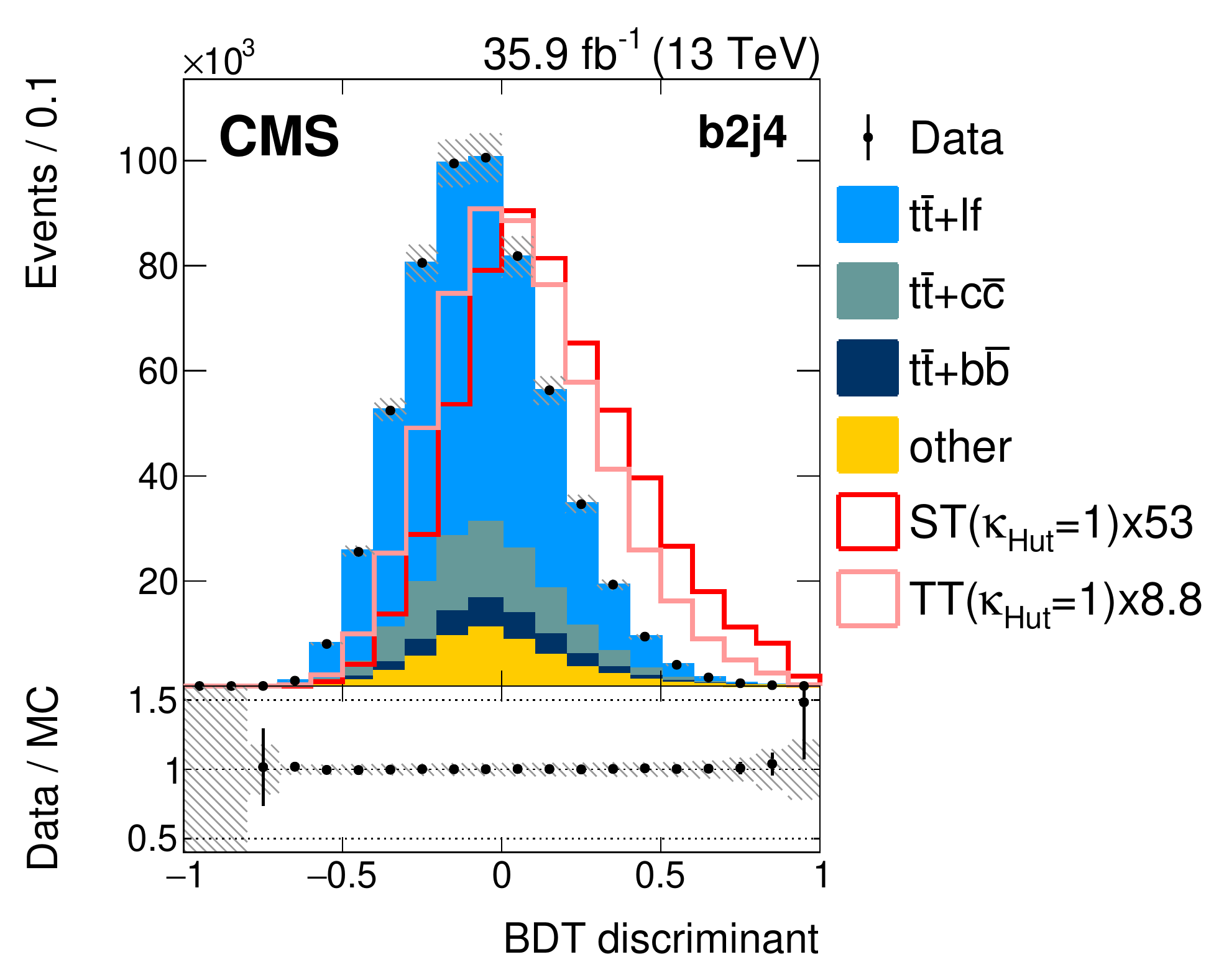}
    \includegraphics[width=0.45\textwidth]{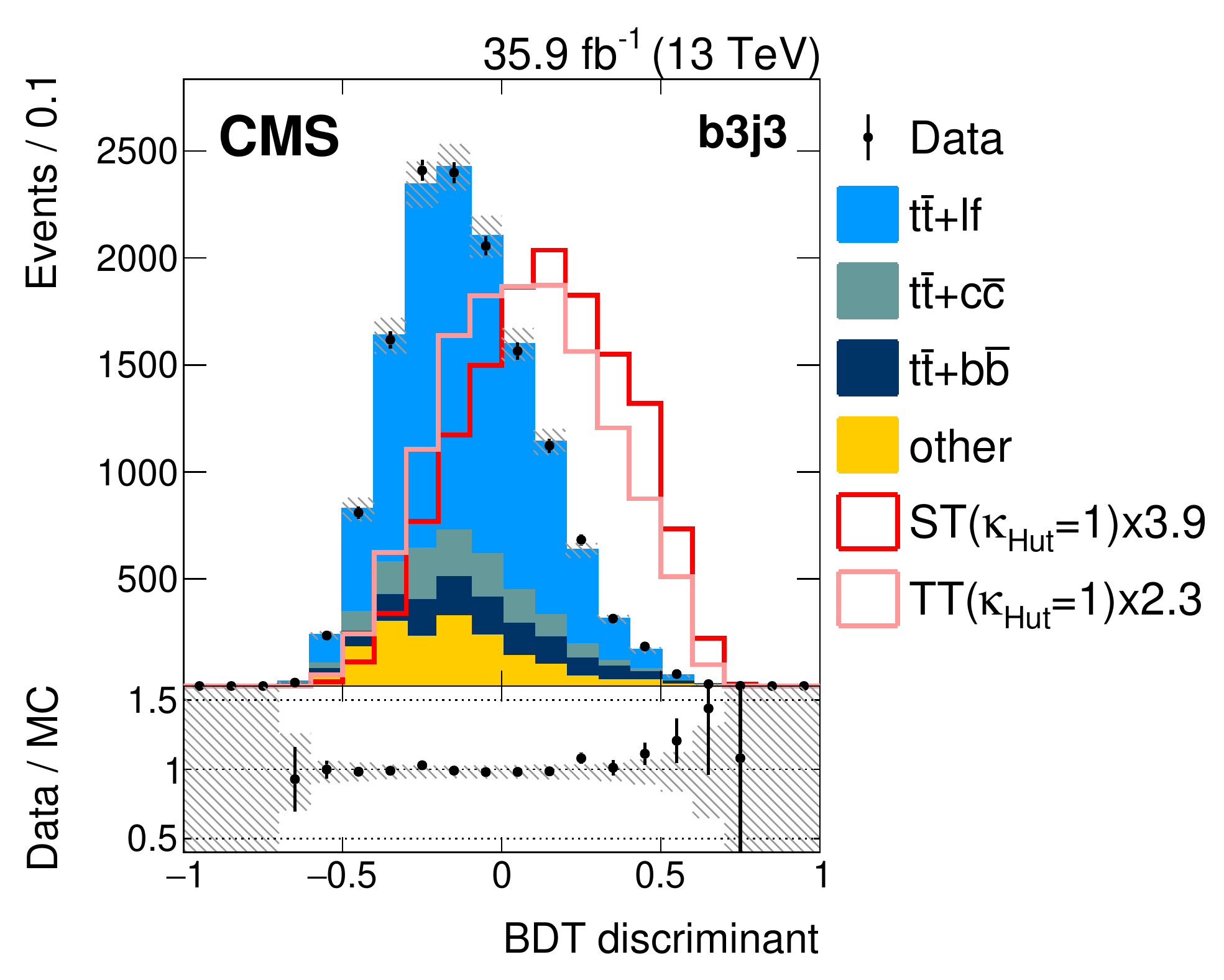}
    \includegraphics[width=0.45\textwidth]{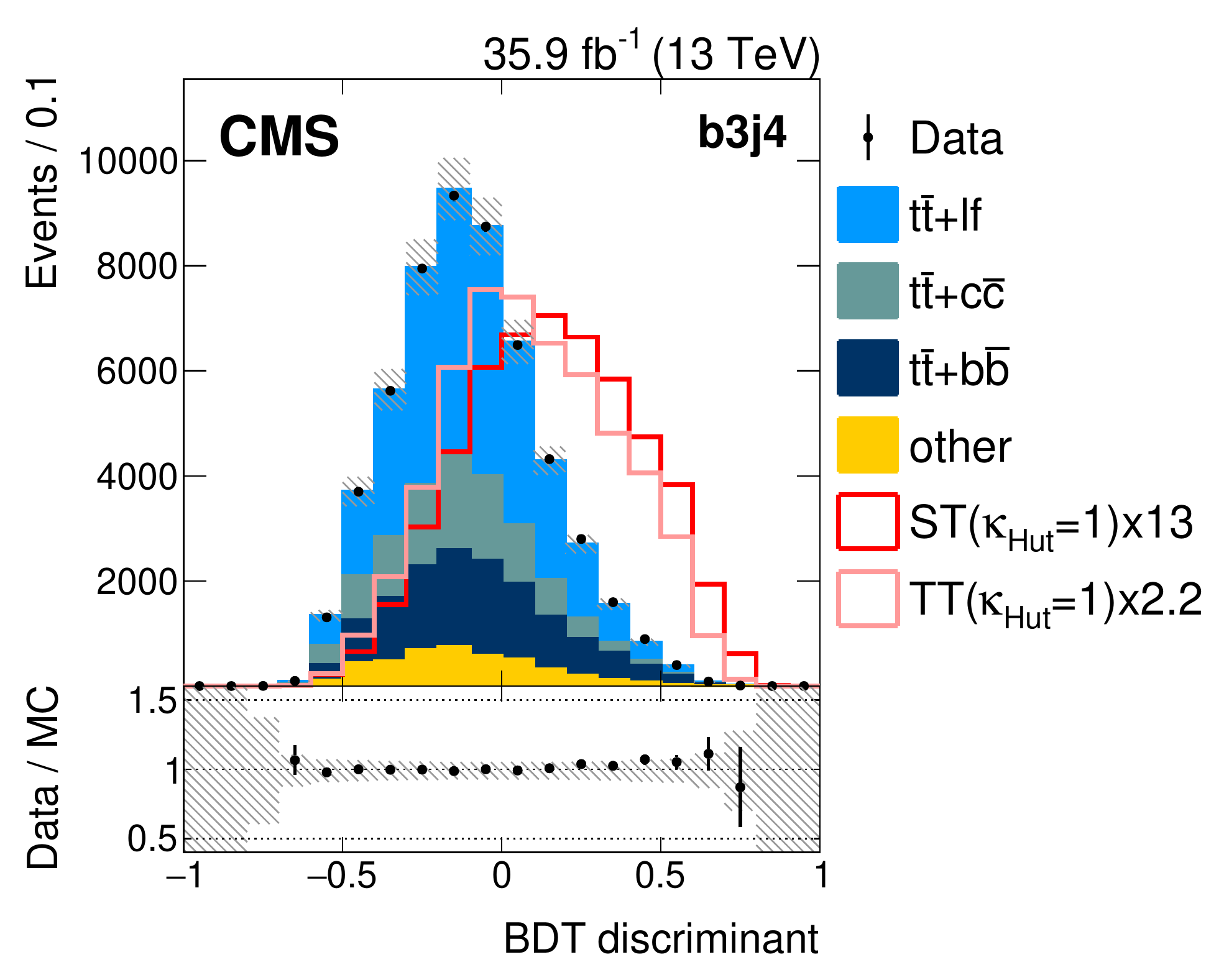}
    \caption{The BDT discriminant distributions for different jet categories for
    Hut training after the fit to data. All background processes are
    constrained to the SM expectation in the fit.
    The shaded area corresponds to the total
    uncertainty in the predicted background.
    The data-to-simulation ratio is also shown. The distributions for
    the signal processes are normalized to the total number of events in
    the predicted background to ease the comparison
    of the shapes of the distributions.}
    \label{fig:BDTHut}
\end{figure}

\begin{figure}[hp]
  \centering
    \includegraphics[width=0.45\textwidth]{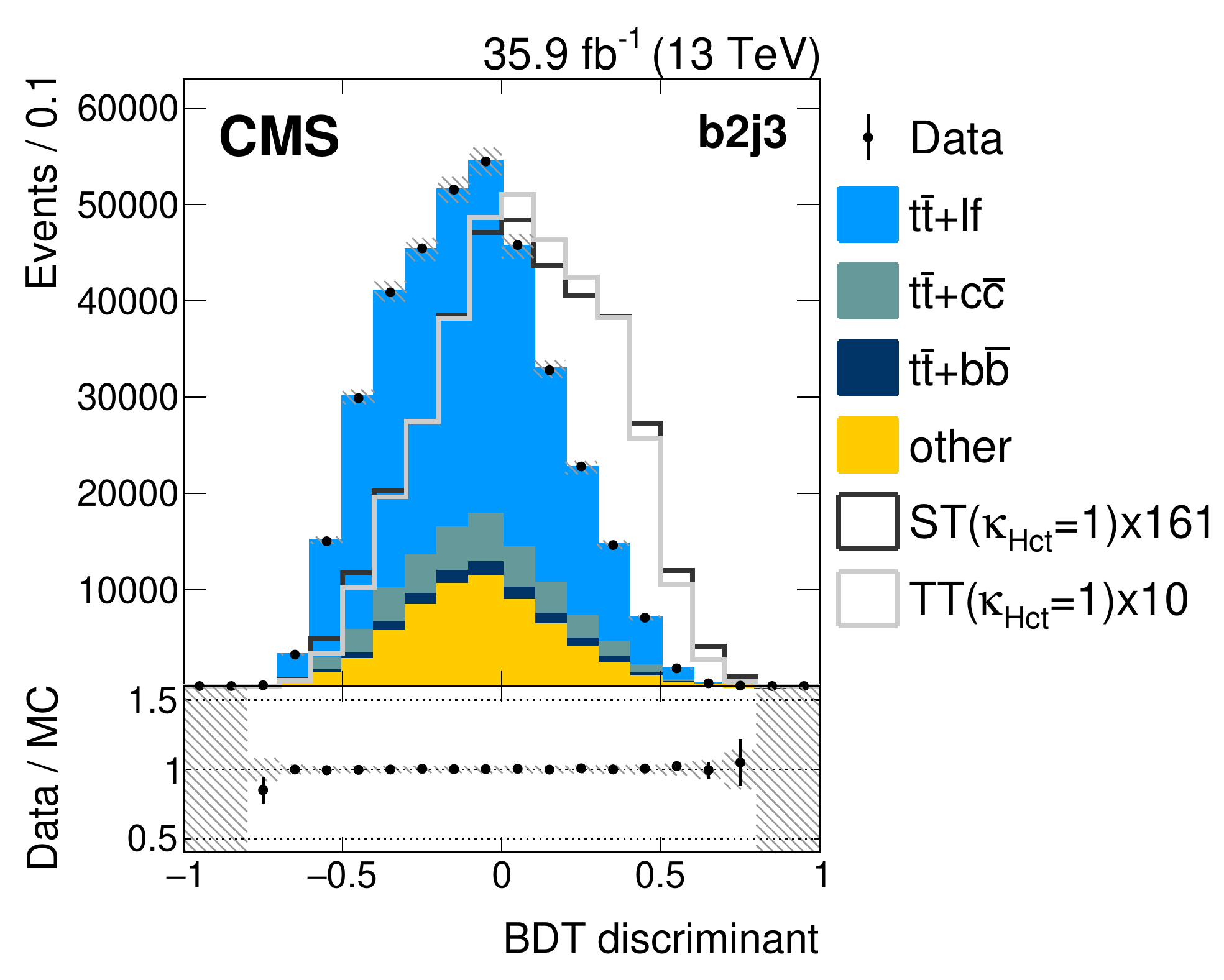}
    \includegraphics[width=0.45\textwidth]{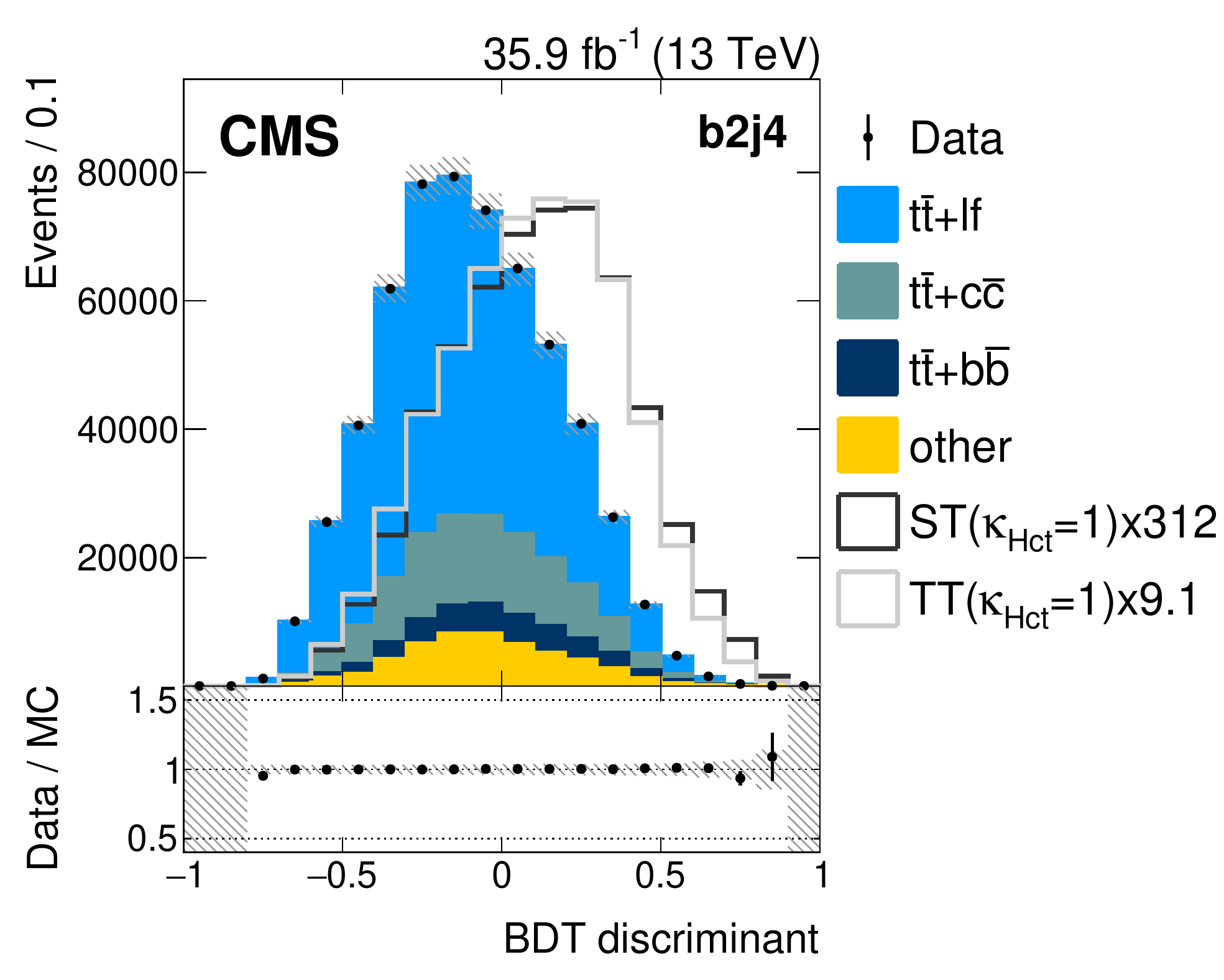}
    \includegraphics[width=0.45\textwidth]{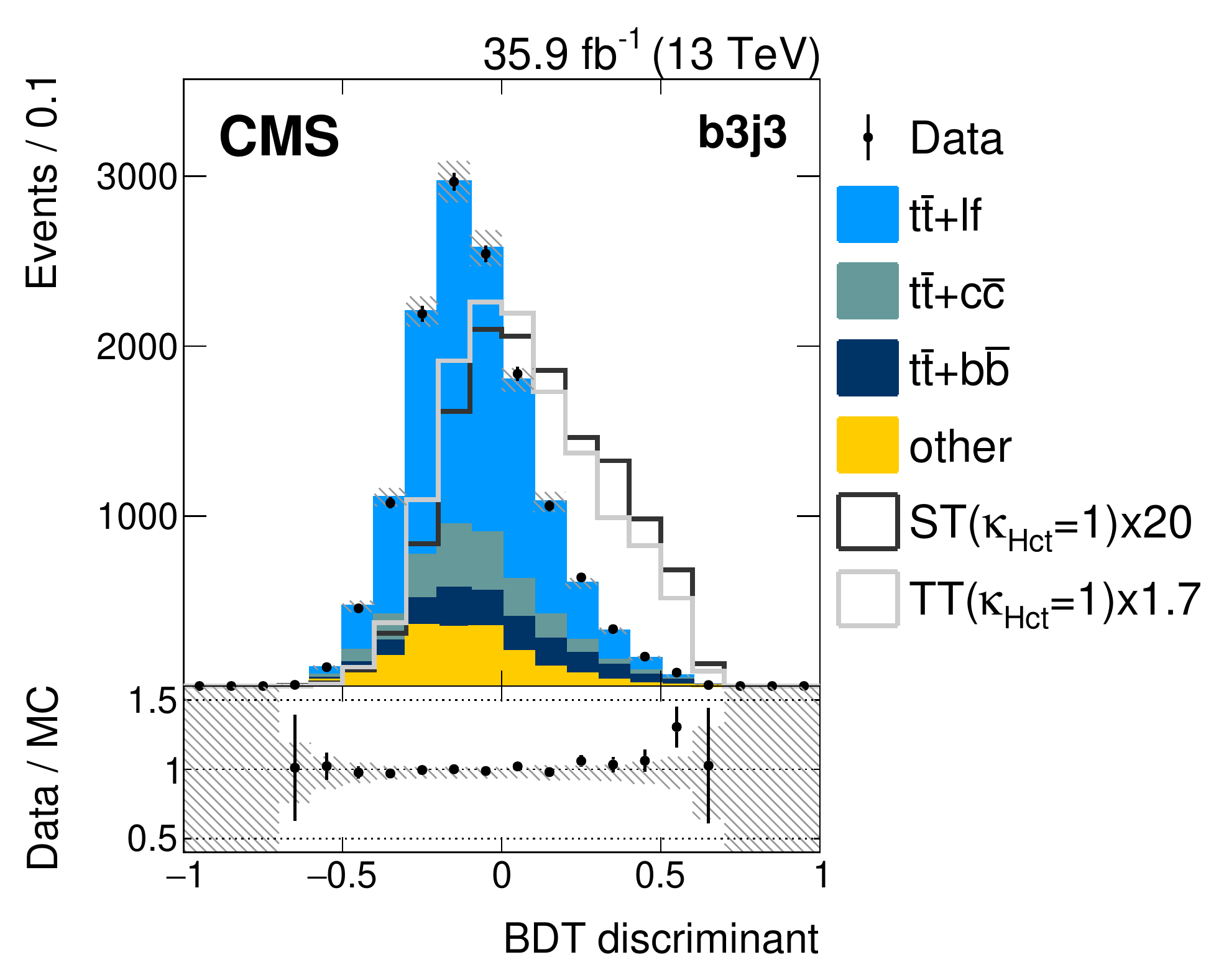}
    \includegraphics[width=0.45\textwidth]{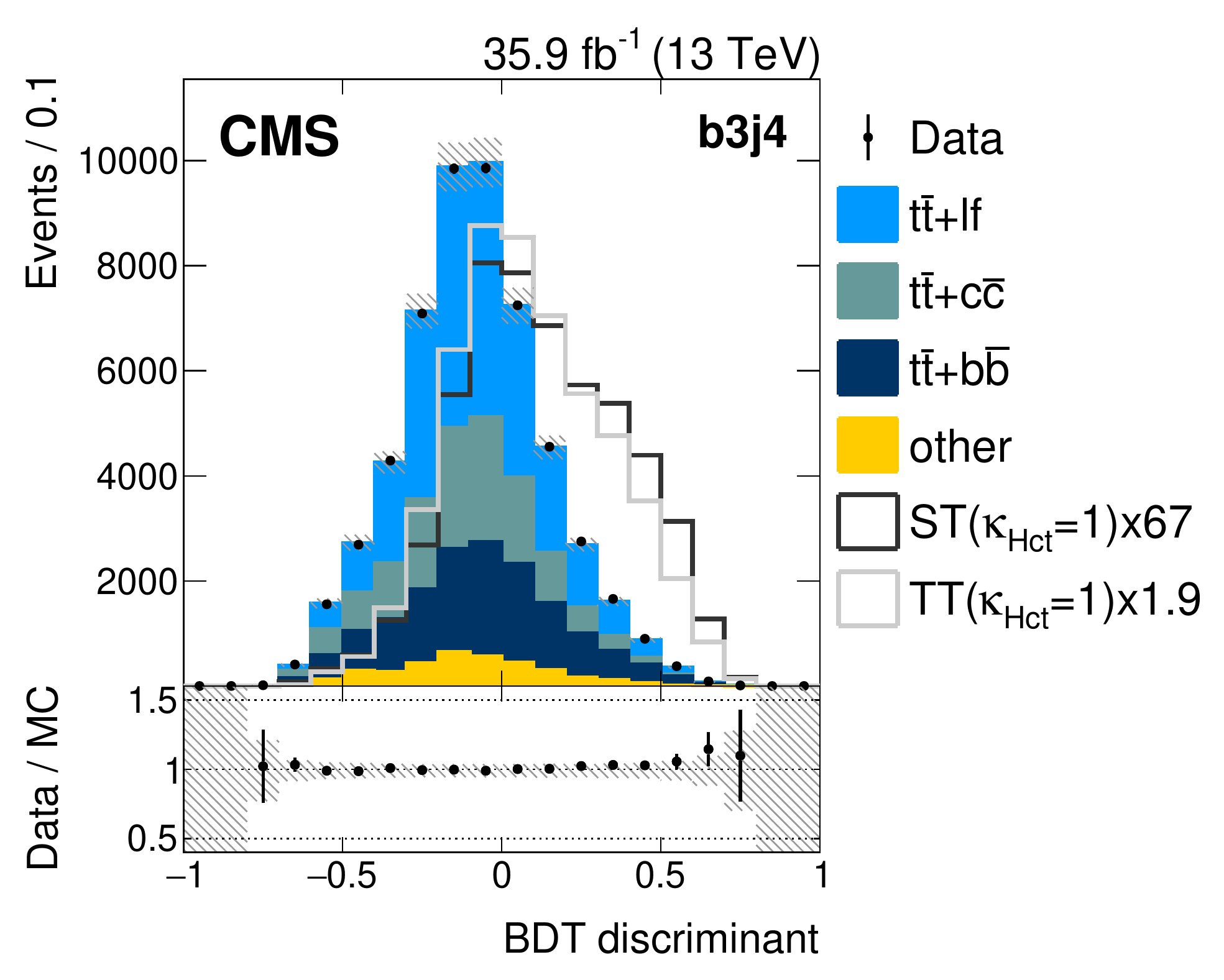}
    \includegraphics[width=0.45\textwidth]{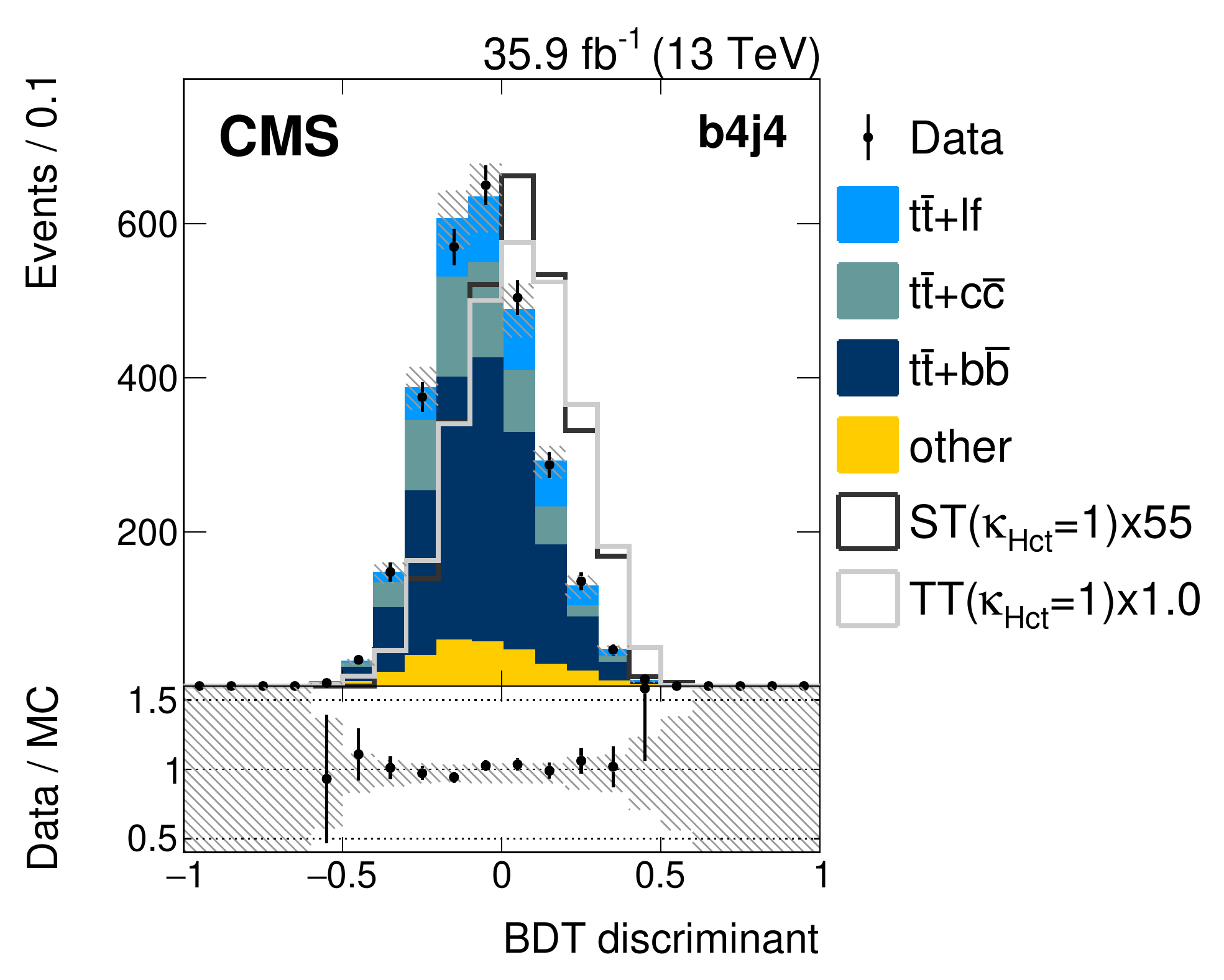}
    \caption{The BDT discriminant distributions for different jet categories for
    Hct training after the fit to data. All background processes are
    constrained to the SM expectation in the fit.
    The shaded area corresponds to the total
    uncertainty in the predicted background.
    The data-to-simulation ratio is also shown. The distributions for
    the signal processes are normalized to the total number of events in
    the predicted background to ease the comparison
    of the shapes of the distributions.}
    \label{fig:BDTHct}
\end{figure}

The final observable used to extract signal events is defined as the BDT score
distribution in each jet category corresponding to either
Hut or Hct signal training. Figures~\ref{fig:BDTHut}
and~\ref{fig:BDTHct} show the comparison between data and simulation for
this observable after the fit to data with all background processes constrained
to the SM expectation.

\section{Estimation of systematic uncertainties}

Sources of systematic uncertainty that affect both the normalization and shape
of the predicted signal and background events are considered in the analysis.
All systematic uncertainties
are treated as nuisance parameters in the derivation of the exclusion
limit.

The dominant systematic uncertainty arises from the application of the b tagging
requirement.
The shape of the CSVv2 discriminant, the b tagging efficiency, and the
misidentification rate in simulation are corrected to reproduce the data
distributions~\cite{CMS-PAS-BTV-15-001}.
The uncertainties that are associated with these correction factors are
the statistical uncertainty due to the limited data sample from which the
correction factors were derived, and the systematic uncertainty arising from the purity estimate
of the sample as predicted by simulation. The overall effect of this systematic uncertainty
leads to a variation of {$\approx$}8--30\% in simulated event yields, with the
largest effect observed in event categories with a large number of b-tagged jets.

The uncertainty associated with the choice of renormalization and
factorization scales in the matrix element is estimated by changing each scale separately by a factor
of 1/2 and 2.
To estimate the systematic uncertainty at the parton-shower level,
several special simulated samples of events are considered, where the scales
used to determine the ISR and FSR emissions are varied.
The uncertainty associated with the choice of PDF is estimated by
using several PDFs and assigning the maximum differences as the quoted
uncertainty, following the PDF4LHC prescription with the
MSTW2008 68\% CL NNLO, CT10 NNLO, and NNPDF2.3 5f
FFN PDF sets (see Ref.~\cite{TT2} and references therein, as well as
Refs.~\cite{TT3,TT4,TT5}).
The overall uncertainty associated with
the simulation of the background processes contributes up to {$\approx$}20\%
in the variation of event yields.

Following the prescription in \POWHEG~\cite{CMS-PAS-TOP-16-021},
the matching of the high-\pt partons, from matrix-element
calculations and parton-shower emission, is regulated by damping the emission by the factor
$m_{\PQt}^2/(p^2_{\mathrm{T}} + m_{\PQt}^2)$.
Additional simulated samples for \ttbar are used that correspond to the variation
of this factor within the considered uncertainty.
For the \ttbar and single top quark $t$-channel simulated samples the
additional systematic uncertainties associated with the amount of
multiparton interactions and color reconnection~\cite{CR1,CR2} are considered.
These uncertainties were determined by varying
them according to the uncertainties reported for the underlying event
tune CUETP8M2T4, and lead to a systematic effect of {$\approx$}1--5\%.

The uncertainty associated with the calibration of the jet energy
scale and the jet energy resolution contributes up to {$\approx$}8\% in
the variation of the final event yields~\cite{JETMET}.
The identification, isolation, and trigger efficiency correction
uncertainties for reconstructed leptons contribute up to 0.5\% of the
total uncertainty in the predicted yield.  An uncertainty of 2.5\% is
assigned to the measured integrated luminosity value of the considered data sample~\cite{lumi}.

The number of simulated pileup events is corrected to match the measured number of events in data. The uncertainty
on the total inelastic cross section is taken as 4.6\%. Its overall
contribution to the total systematic uncertainty is found to be negligible.

The \pt spectrum of individual top quarks in data is found to be softer than
predicted by the simulation.
A correction for the top quark \pt spectrum in simulation is applied
and the difference between the initial and the corrected shapes is taken
as an additional systematic uncertainty~\cite{TOP-16-008}. This uncertainty also has
a negligible impact on the final distributions.

Additionally, a systematic uncertainty of 50\% in the predicted
cross sections for
\ttbar{}+\bbbar
and
\ttbar{}+\ccbar
processes is assumed~\cite{ttbbXsec,ttbbXsec2}.

\section{Results}

{\renewcommand{\arraystretch}{1.3}

A comparison between the number of selected events in data and
simulation is shown in Tables~\ref{tab:eventcountsHut}
and~\ref{tab:eventcountsHct}.
A 95\% CL upper limit is computed for the production cross section of
$\PQt\PH$ FCNC events times branching fractions of top quark semileptonic
decay and Higgs boson decay to $\PQb$ quarks
that uses the asymptotic approximation of the $\mathrm{CL_{s}}$
method~\cite{ARead,TJunk}. The profile likelihood ratio test
statistic~\cite{GCowan} is defined as $q(\mu) =
-2\ln(L(\mu,\hat{\theta}_{\mu})/L(\hat{\mu},\hat{\theta}))$, where $L$ is a
binned likelihood function, $\mu$ is a signal strength modifier,
$\theta$ is a set of nuisance parameters, $\hat{\theta}_{\mu}$ is a
set of nuisance parameters that maximize $L$ for a given $\mu$, $\hat{\mu}$ and
$\hat{\theta}$ are the values of the corresponding parameters which
simultaneously maximize $L$.
Uncertainties due to normalization are included through nuisance parameters with log-normal prior
distributions, while shape uncertainties are included with Gaussian
prior distributions. The expected and observed $95\%$ CL upper limits are derived on the
signal production cross section separately for each event category, as
well as for their combination (Fig.~\ref{fig:Limit}). In the
latter case, a simultaneous binned maximum-likelihood fit to all
categories is performed. The fit takes into account the statistical
and systematic uncertainties in the final BDT score distributions
in each jet category.

\begin{table}
  \centering
    \topcaption{Number of events in each category together with its
    total relative uncertainty as obtained from the fit to data for
    Hut.}
    \begin{tabular}{c|cccc}
   & b2j3 & b2j4 & b3j3 & b3j4 \\
  \hline
  Data & 365 890 & 575 500 & 13 481 & 53 352 \\
   \hline
   \ttbar{}+\bbbar
& 8 880 $\pm$ 3 641 & 30 157 $\pm$ 5 127 & 1 214 $\pm$ 510 & 11 668 $\pm$ 1 750 \\
\ttbar{}+\ccbar
& 26 035 $\pm$ 11 195 & 81 959 $\pm$ 18 031 & 1 281 $\pm$ 576 & 9 753 $\pm$ 2 243 \\
   \ttbar{}+lf & 270 989 $\pm$ 13 820 & 410 028 $\pm$ 16 401 & 9 104 $\pm$ 674 & 27 079 $\pm$ 1 733 \\
   other & 58 991 $\pm$ 6 489 & 51 845 $\pm$ 6 221 & 1 616 $\pm$ 356 &
   4 269 $\pm$ 768 \\ \hline
   Total & 364 895 $\pm$ 22 623 & 573 989 $\pm$ 25 256 & 13 215 $\pm$ 1 255 & 52 769 $\pm$ 3 430 \\
  \end{tabular}
  \label{tab:eventcountsHut}
\end{table}

\begin{table}
  \centering
    \topcaption{Number of events in each category together with its
    total relative uncertainty as obtained from the fit to data for
    Hct.}
    \begin{tabular}{c|ccccc}
   & b2j3 & b2j4 & b3j3 & b3j4 & b4j4 \\
  \hline
  Data & 365 890 & 575 500 & 13 481 & 53 352 & 2 764 \\
  \hline
  \ttbar{}+\bbbar
& 10 176 $\pm$ 1 933 & 34 174 $\pm$ 3 759 & 1 367 $\pm$ 273 & 12 897 $\pm$ 1 058 & 1 517 $\pm$ 129 \\
\ttbar{}+\ccbar
& 33 210 $\pm$ 11 956 & 102 186 $\pm$ 15 328 & 1 674 $\pm$ 619 & 12 280 $\pm$ 1 842 & 521 $\pm$ 104 \\
   \ttbar{}+lf & 258 679 $\pm$ 8 795 & 385 395 $\pm$ 10 791 & 8 349 $\pm$ 451 & 24 083 $\pm$ 1 132 & 383 $\pm$ 69 \\
   other & 62 887 $\pm$ 5 723 & 52 134 $\pm$ 6 256 & 1 742 $\pm$ 401 &
   3 513 $\pm$ 562 & 262 $\pm$ 50 \\ \hline
   Total & 364 952 $\pm$ 16 788 & 573 889 $\pm$ 18 364 & 13 132 $\pm$ 959 & 52 773 $\pm$ 2 322 & 2 682 $\pm$ 185 \\
  \end{tabular}
  \label{tab:eventcountsHct}
\end{table}

\begin{figure}
  \centering
    \includegraphics[width=0.45\textwidth]{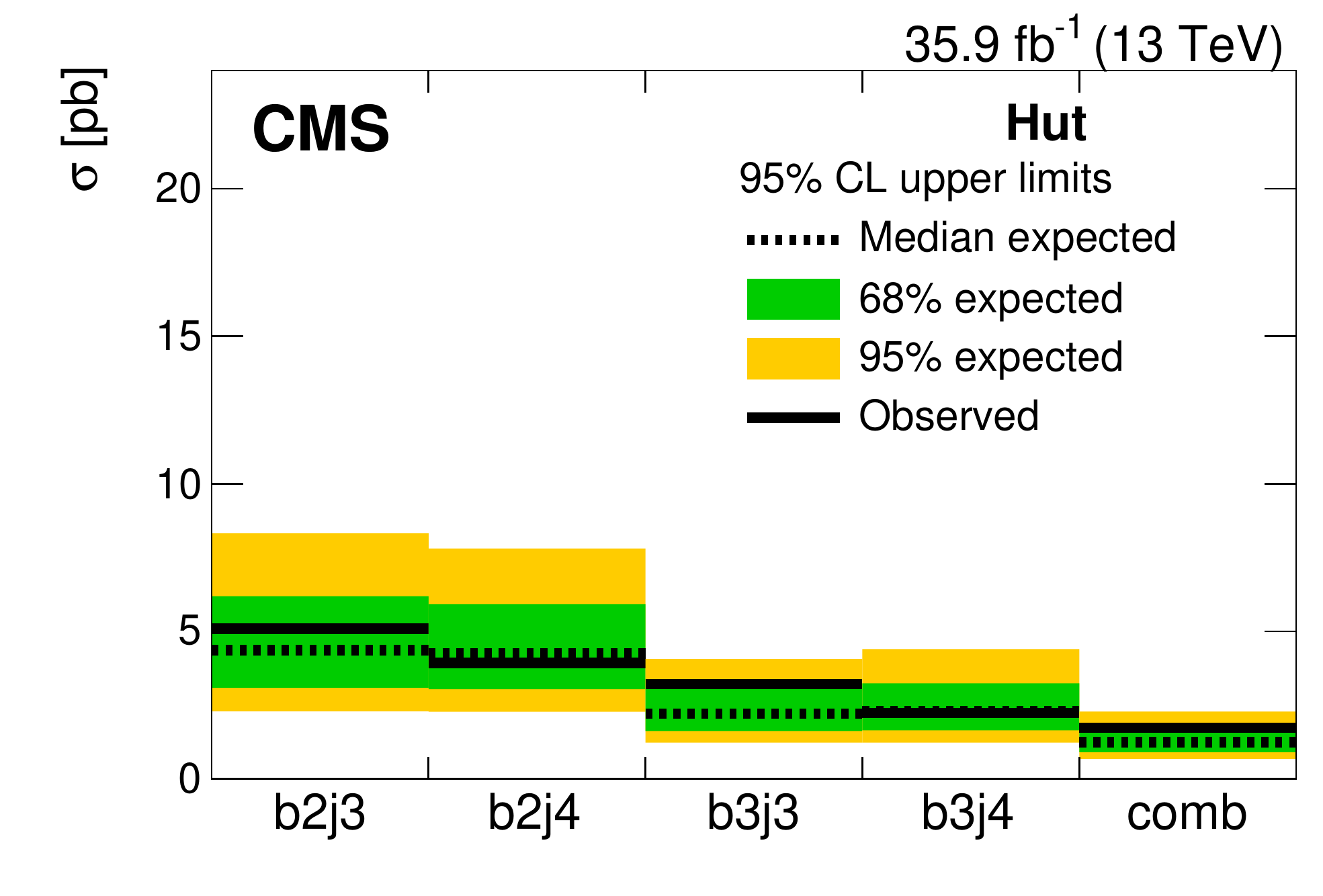}
    \includegraphics[width=0.45\textwidth]{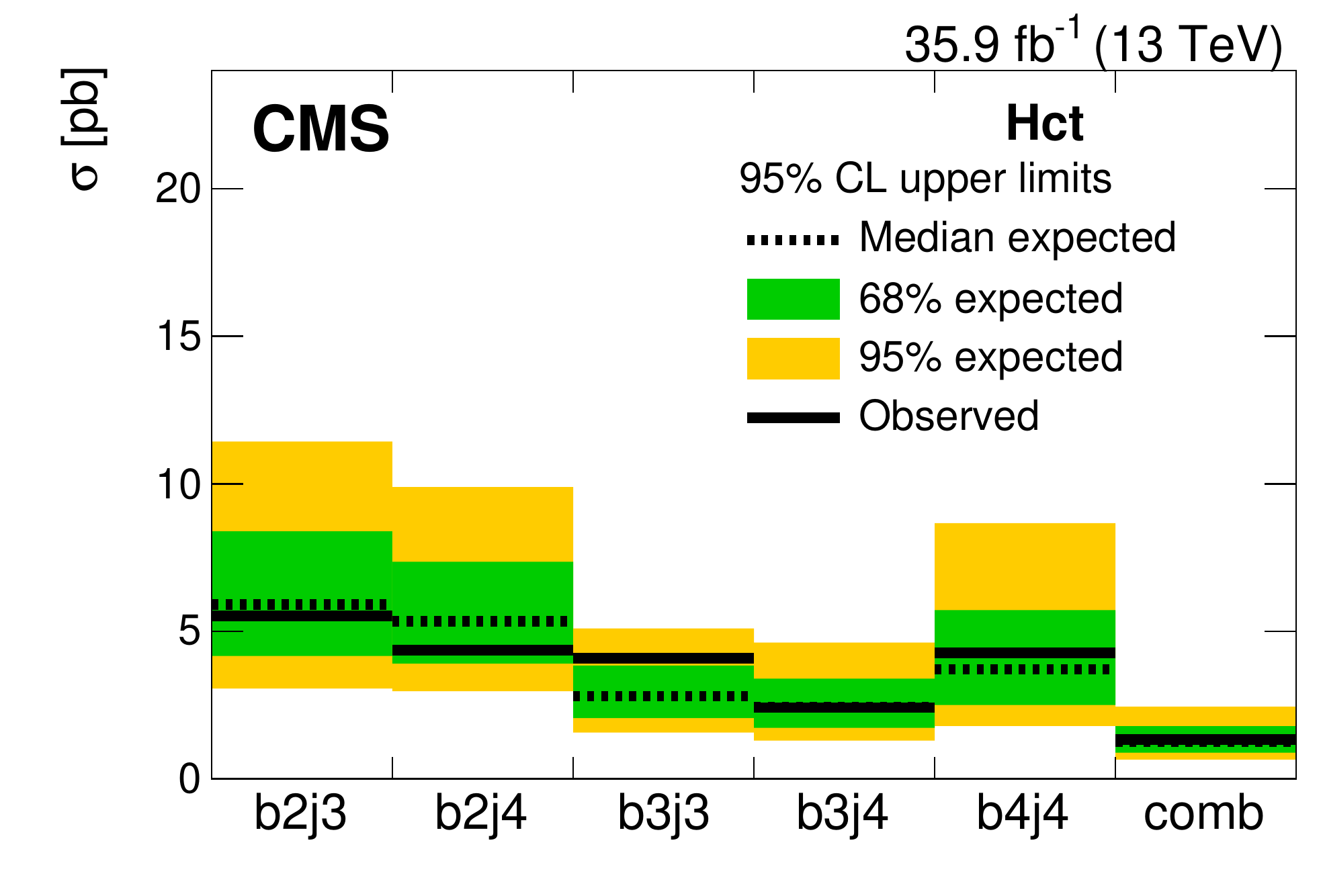}
    \caption{Excluded signal cross section at 95\% CL per event category for
    Hut (left) and Hct (right).}
    \label{fig:Limit}
\end{figure}

\begin{figure}
  \centering
    \includegraphics[width=0.6\textwidth]{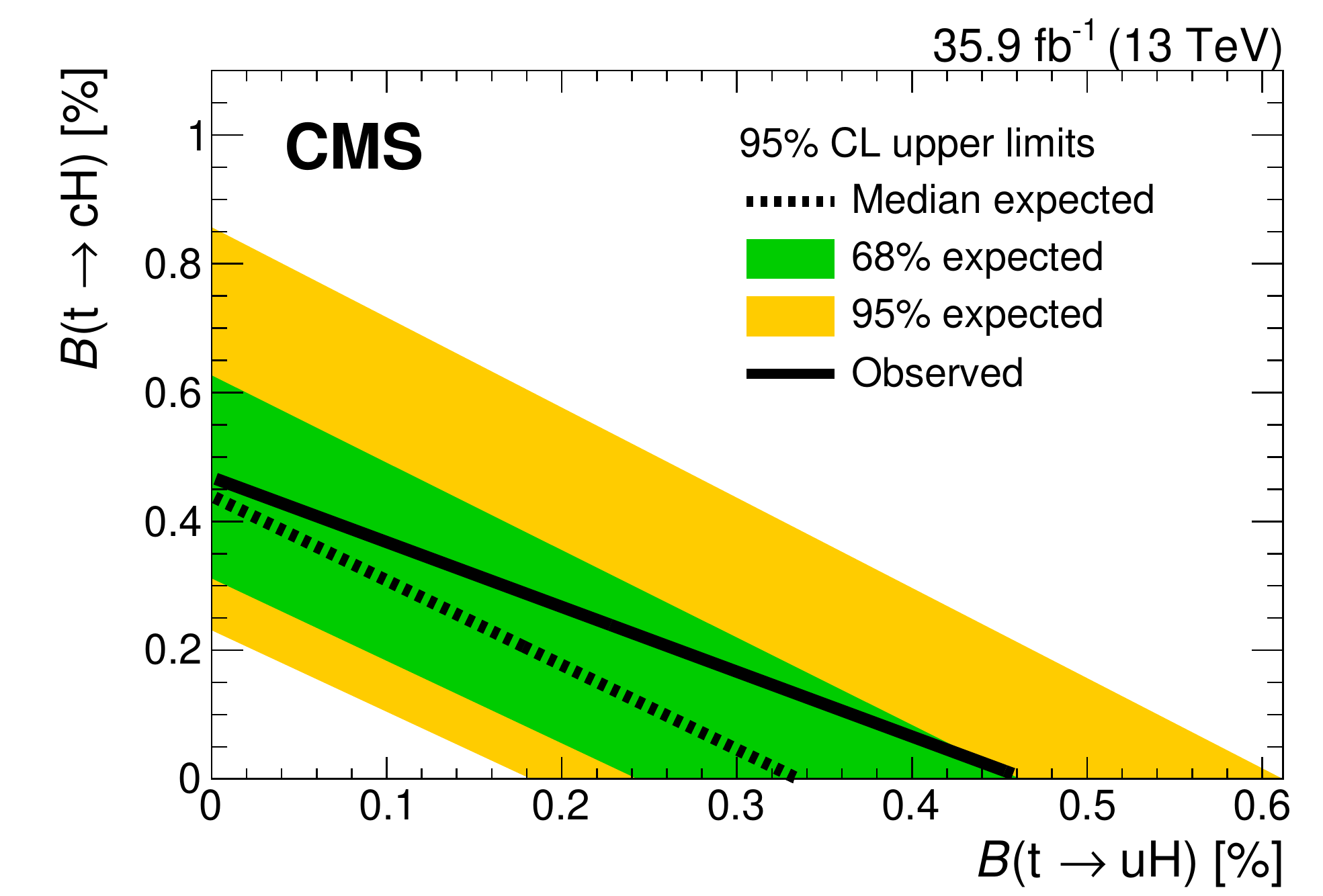}
    \caption{Upper limits on $\mathcal{B}(\PQt \to \PQu\PH)$
    and $\mathcal{B}(\PQt \to \PQc\PH)$ at 95\% CL.}
    \label{fig:Limit2D}
\end{figure}

\begin{figure}
  \centering
    \includegraphics[width=0.6\textwidth]{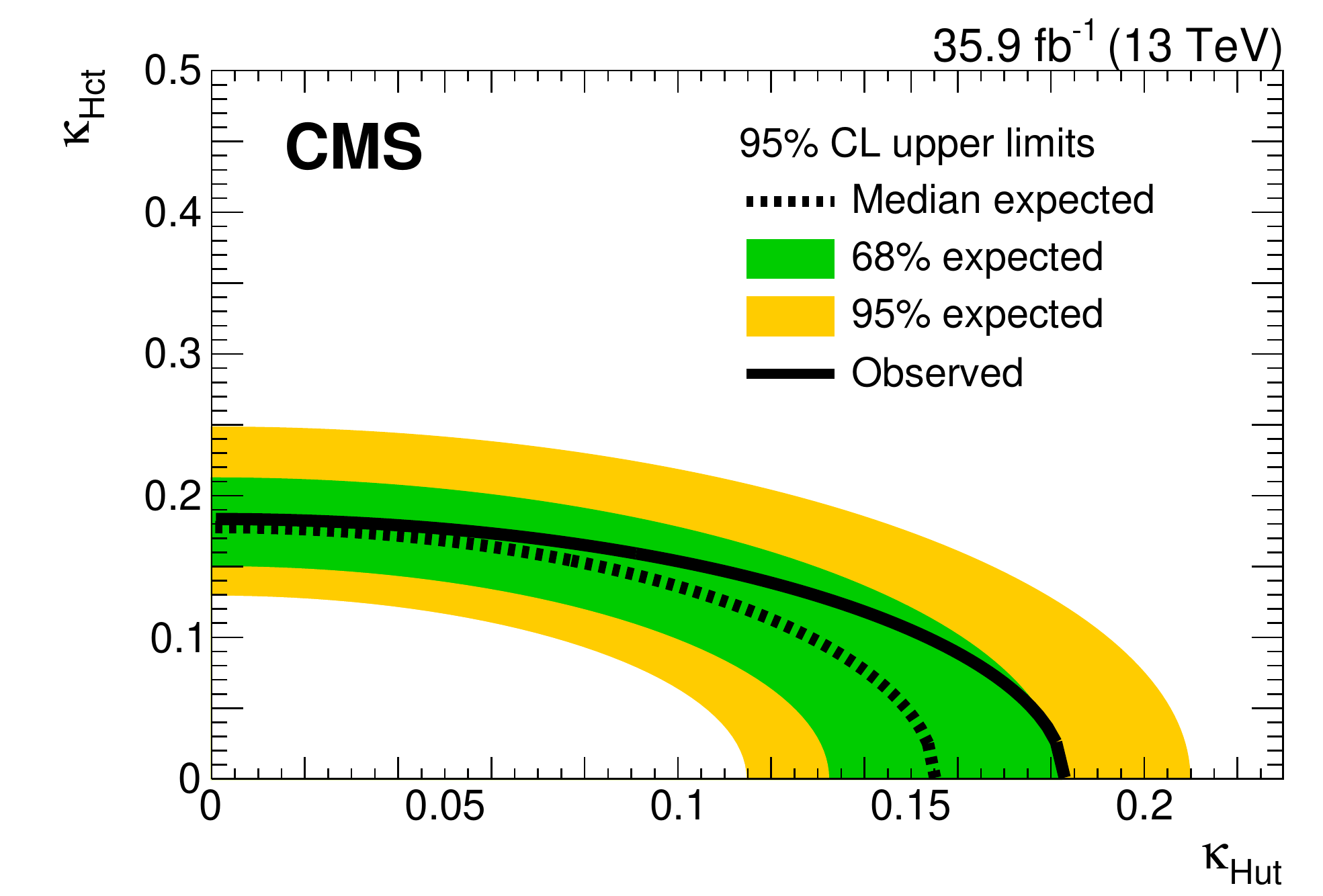}
    \caption{Upper limits on $\kappa_{\PH\PQu\PQt}$ and
    $\kappa_{\PH\PQc\PQt}$ at 95\% CL.}
    \label{fig:Limit2DCoup}
\end{figure}

\begin{figure}
  \centering
    \includegraphics[width=0.45\textwidth]{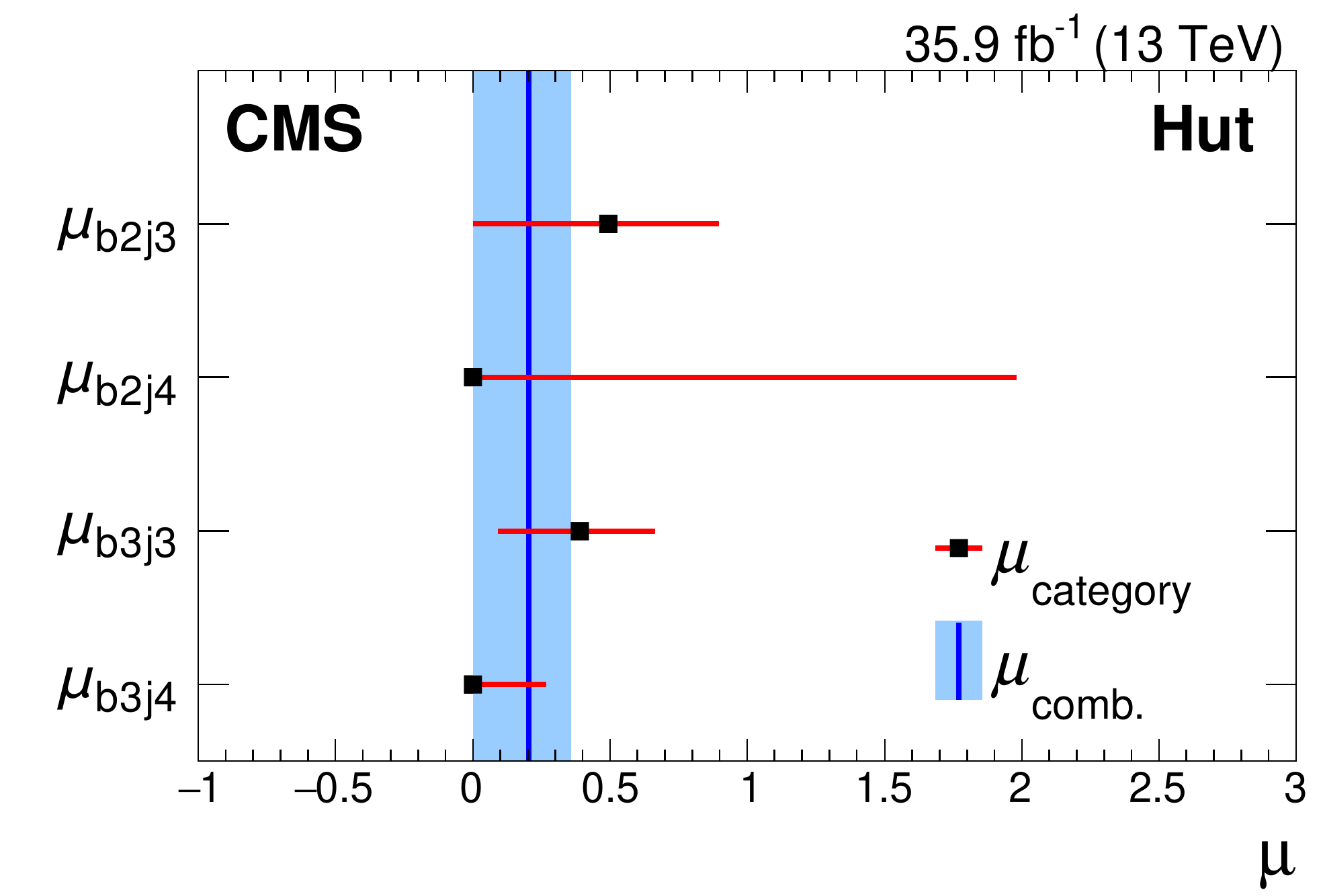}
    \includegraphics[width=0.45\textwidth]{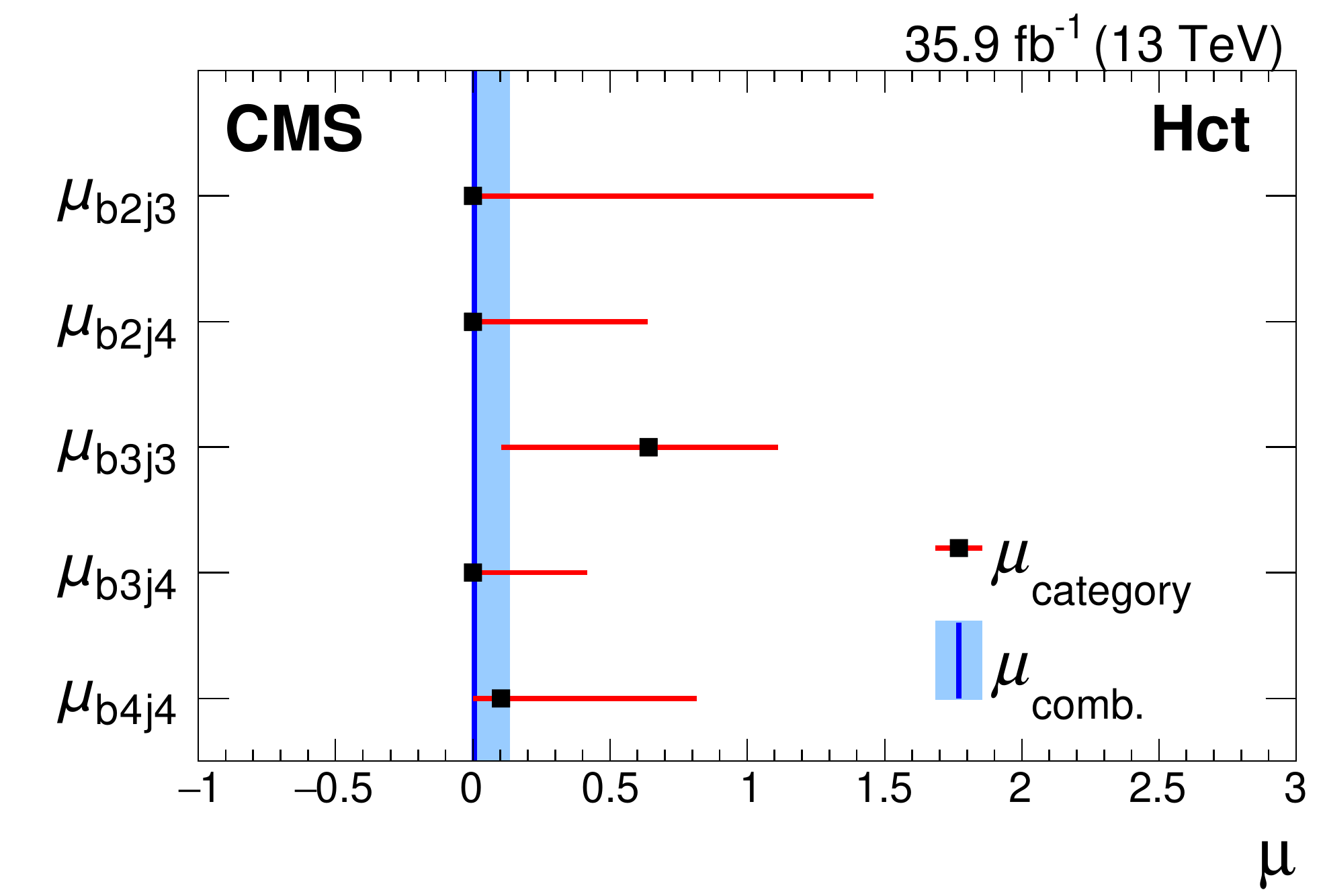}
    \caption{The best fit signal strength ($\mu$) for Hut (left) and Hct
    (right), which is restricted to positive values in the fit.}
    \label{fig:BestFit}
\end{figure}

The resultant observed (expected) 95\% CL exclusion limits on top quark FCNC decay branching
fractions are $\mathcal{B}(\PQt \to \PQu\PH) < 0.47\%\,(0.34\%)$ and
$\mathcal{B}(\PQt \to \PQc\PH) <0.47\%\,(0.44\%)$.
These upper limits on the branching fractions can be translated into upper
limits on the coupling strengths using the relations:
\begin{equation}
\begin{aligned}
\kappa_{\PH\PQu\PQt}^2 &= \mathcal{B}(\PQt \to
\PQu\PH)\frac{\Gamma_{\PQt}}{\Gamma_{\PH\PQu\PQt}},\\
\kappa_{\PH\PQc\PQt}^2 &= \mathcal{B}(\PQt \to
\PQc\PH)\frac{\Gamma_{\PQt}}{\Gamma_{\PH\PQc\PQt}},
\end{aligned}
\end{equation}
where the total top quark width is $\Gamma_{\PQt} =
1.32$\GeV~\cite{TopWidth},
and the partial width for the FCNC decay process of the top quark is
$\Gamma_{\PH\PQu\PQt} = \Gamma_{\PH\PQc\PQt} = 0.184$\GeV for
$\kappa_{\PH\PQu\PQt} = \kappa_{\PH\PQc\PQt} = 1$.
The resultant limits on the coupling strengths are $\kappa_{\PH\PQu\PQt} < 0.18\,(0.16)$ and
$\kappa_{\PH\PQc\PQt} < 0.18\,(0.18)$.
These limits are very competitive to the CMS result with the combination of various channels at 8
TeV~\cite{TopHiggsCMS},
while the ATLAS result with the analysis of the $\PH \to
\PGg\PGg$ decay at 13\TeV~\cite{TopHiggsATLAS13TeV} represents the best limits to date.
The measured one-dimensional exclusion limits are
also interpreted for the scenario of the non-vanishing
FCNC couplings via a linear interpolation.
The results for two-dimensional limits on top quark FCNC decay
branching fractions and coupling strengths are presented in
Fig.~\ref{fig:Limit2D} and~\ref{fig:Limit2DCoup}, respectively. We define a signal
strength parameter as $\mu = \sigma/\sigma_{\text{sig}}$, where $\sigma$ is the
cross section excluded at $95\%$ CL and $\sigma_{\text{sig}}$ is the predicted
cross section for signal.
A maximum likelihood fit is performed for the signal strength, and is
shown in Fig.~\ref{fig:BestFit}.
Inclusion of the associated production of a single top quark with
a Higgs boson in this study provides a {$\approx$}20\% relative improvement in the
expected upper limit on $\mathcal{B}(\PQt \to \PQu\PH)$ with
respect to the results obtained in an analysis of only \ttbar events with top quark FCNC decays.

\section{Summary}

A search for flavor-changing neutral currents in events with a top
quark and the Higgs boson, corresponding to a data sample of
35.9\fbinv collected in proton-proton collisions at
$\sqrt{s}=13\TeV$, is presented. This is the first search to probe
$\PQt\PH$ flavor-changing neutral current couplings in both associated
production of a top quark
with the Higgs boson and in top quark decays.
Observed (expected) upper limits at 95\% confidence level are
set on the branching fractions of top quark decays, $\mathcal{B}(\PQt \to \PQu\PH) <0.47\%\,(0.34\%)$
and $\mathcal{B}(\PQt \to \PQc\PH) <0.47\%\,(0.44\%)$.
These results provide a significant improvement over the previous
limits set by CMS in the $\PH \to \bbbar$ channel, as well as
represent the best limits for $\mathcal{B}(\PQt
\to \PQu\PH)$ at CMS.

\begin{acknowledgments}
We congratulate our colleagues in the CERN accelerator departments for the excellent performance of the LHC and thank the technical and administrative staffs at CERN and at other CMS institutes for their contributions to the success of the CMS effort. In addition, we gratefully acknowledge the computing centres and personnel of the Worldwide LHC Computing Grid for delivering so effectively the computing infrastructure essential to our analyses. Finally, we acknowledge the enduring support for the construction and operation of the LHC and the CMS detector provided by the following funding agencies: BMWFW and FWF (Austria); FNRS and FWO (Belgium); CNPq, CAPES, FAPERJ, and FAPESP (Brazil); MES (Bulgaria); CERN; CAS, MoST, and NSFC (China); COLCIENCIAS (Colombia); MSES and CSF (Croatia); RPF (Cyprus); SENESCYT (Ecuador); MoER, ERC IUT, and ERDF (Estonia); Academy of Finland, MEC, and HIP (Finland); CEA and CNRS/IN2P3 (France); BMBF, DFG, and HGF (Germany); GSRT (Greece); OTKA and NIH (Hungary); DAE and DST (India); IPM (Iran); SFI (Ireland); INFN (Italy); MSIP and NRF (Republic of Korea); LAS (Lithuania); MOE and UM (Malaysia); BUAP, CINVESTAV, CONACYT, LNS, SEP, and UASLP-FAI (Mexico); MBIE (New Zealand); PAEC (Pakistan); MSHE and NSC (Poland); FCT (Portugal); JINR (Dubna); MON, RosAtom, RAS, RFBR and RAEP (Russia); MESTD (Serbia); SEIDI, CPAN, PCTI and FEDER (Spain); Swiss Funding Agencies (Switzerland); MST (Taipei); ThEPCenter, IPST, STAR, and NSTDA (Thailand); TUBITAK and TAEK (Turkey); NASU and SFFR (Ukraine); STFC (United Kingdom); DOE and NSF (USA).

\hyphenation{Rachada-pisek} Individuals have received support from the Marie-Curie programme and the European Research Council and Horizon 2020 Grant, contract No. 675440 (European Union); the Leventis Foundation; the A. P. Sloan Foundation; the Alexander von Humboldt Foundation; the Belgian Federal Science Policy Office; the Fonds pour la Formation \`a la Recherche dans l'Industrie et dans l'Agriculture (FRIA-Belgium); the Agentschap voor Innovatie door Wetenschap en Technologie (IWT-Belgium); the Ministry of Education, Youth and Sports (MEYS) of the Czech Republic; the Council of Science and Industrial Research, India; the HOMING PLUS programme of the Foundation for Polish Science, cofinanced from European Union, Regional Development Fund, the Mobility Plus programme of the Ministry of Science and Higher Education, the National Science Center (Poland), contracts Harmonia 2014/14/M/ST2/00428, Opus 2014/13/B/ST2/02543, 2014/15/B/ST2/03998, and 2015/19/B/ST2/02861, Sonata-bis 2012/07/E/ST2/01406; the National Priorities Research Program by Qatar National Research Fund; the Programa Severo Ochoa del Principado de Asturias; the Thalis and Aristeia programmes cofinanced by EU-ESF and the Greek NSRF; the Rachadapisek Sompot Fund for Postdoctoral Fellowship, Chulalongkorn University and the Chulalongkorn Academic into Its 2nd Century Project Advancement Project (Thailand); the Welch Foundation, contract C-1845; and the Weston Havens Foundation (USA).
\end{acknowledgments}
\bibliography{auto_generated}

\cleardoublepage \appendix\section{The CMS Collaboration \label{app:collab}}\begin{sloppypar}\hyphenpenalty=5000\widowpenalty=500\clubpenalty=5000\input{TOP-17-003-authorlist.tex}\end{sloppypar}
\end{document}

%% file: TOP-17-003-authorlist.tex
\vskip\cmsinstskip
\textbf{Yerevan Physics Institute,  Yerevan,  Armenia}\\*[0pt]
A.M.~Sirunyan,  A.~Tumasyan
\vskip\cmsinstskip
\textbf{Institut f\"{u}r Hochenergiephysik,  Wien,  Austria}\\*[0pt]
W.~Adam,  F.~Ambrogi,  E.~Asilar,  T.~Bergauer,  J.~Brandstetter,  E.~Brondolin,  M.~Dragicevic,  J.~Er\"{o},  A.~Escalante Del Valle,  M.~Flechl,  M.~Friedl,  R.~Fr\"{u}hwirth\cmsAuthorMark{1},  V.M.~Ghete,  J.~Grossmann,  J.~Hrubec,  M.~Jeitler\cmsAuthorMark{1},  A.~K\"{o}nig,  N.~Krammer,  I.~Kr\"{a}tschmer,  D.~Liko,  T.~Madlener,  I.~Mikulec,  E.~Pree,  N.~Rad,  H.~Rohringer,  J.~Schieck\cmsAuthorMark{1},  R.~Sch\"{o}fbeck,  M.~Spanring,  D.~Spitzbart,  A.~Taurok,  W.~Waltenberger,  J.~Wittmann,  C.-E.~Wulz\cmsAuthorMark{1},  M.~Zarucki
% ...
\vskip\cmsinstskip
% ...
1:~Also at Vienna University of Technology,  Vienna,  Austria\\
% ...
